\documentclass[12pt]{article}

\usepackage[totalheight = 23cm, totalwidth = 17cm]{geometry}
\usepackage{amssymb,amsmath,amsfonts,amsbsy,graphicx,bm}
\usepackage{color,cancel,ulem,url}
\usepackage{axodraw2}

\newcommand{\mpl}{m_{\rm Pl}}
\newcommand{\calA}{{\cal A}}

\newcommand{\calR}{{\cal R}}

\begin{document}

\begin{titlepage}

\begin{center}

{\LARGE \bf 
Instability of  de Sitter space
\\ \vspace{0.25em}
under thermal radiation in different vacua
}

\vskip 1.0cm

{\large
Jinn-Ouk Gong$^{a,b}$ 
and 
Min-Seok Seo$^{c}$ 
}

\vskip 0.5cm

{\it
$^{a}$Department of Science Education, Ewha Womans University
\\ 
Seoul 03760, Korea
\\
$^{b}$Asia Pacific Center for Theoretical Physics, Pohang 37673, Korea
\\
$^{c}$Department of Physics Education, Korea National University of Education
\\ 
Cheongju 28173, Korea
}

\vskip 1.2cm

\end{center}

\begin{abstract}

We study the instability of de Sitter space-time (dS) under thermal radiation in different vacua. 
We argue that  the mode function solution of a scalar field in four-dimensional dS can be separated into the incoming and outgoing modes. Different vacua for dS are realized by different combinations of positive frequency modes assigned to each solution. For a minimally coupled massless scalar field, we explicitly compute the behavior of the mode function and the corresponding energy-momentum tensor in the Unruh vacuum near the horizon, and find that the horizon area increases (decreases) in time when the incoming (outgoing) mode contributes to thermal flux.

\end{abstract}

\end{titlepage}

\newpage

\section{Introduction}

The observation that the semi-classical description of space-time with horizon follows thermodynamic laws leads to the remarkable prediction for black hole physics: Thermal radiation emitted from the event horizon backreacts on the geometry and eventually induces black hole evaporation~\cite{Hawking:1974rv, Hawking:1974sw}. Whereas the final stage of an evaporating black hole is still paradoxical due to our ignorance on quantum gravity at the extremely short length scale~\cite{Hawking:1976ra} (for a review on recent developments, see e.g.,~\cite{Almheiri:2020cfm}) this is an example of the instability of the classical background as a solution to the Einstein equation under quantum corrections. Since de Sitter space-time (dS) also possesses thermodynamic description associated with the event horizon~\cite{Gibbons:1977mu}, it is suggestive to investigate whether dS is stable in the presence of thermal radiation  (see e.g., \cite{Akhmedov:2020ryq} for a recent discussion). Indeed, the breaking of the SO(1,4) dS isometry through the interaction between matter fields and gravity has been a long-standing issue (see e.g.,~\cite{Ford:1984hs, Mottola:1985qt, Frolov:2002va, Polyakov:2007mm, Anderson:2013zia, Rajaraman:2016nvv, Markkanen:2016jhg, Markkanen:2016vrp, Markkanen:2017abw, Aalsma:2019rpt}). Recently, this issue has been revisited under the name of the ``swampland conjectures'', namely, the dS swampland conjecture~\cite{Obied:2018sgi} (see also~\cite{Andriot:2018wzk, Garg:2018reu, Danielsson:2018qpa} and especially~\cite{Ooguri:2018wrx} for the justification based on thermodynamic arguments) and the trans-Planckian censorship conjecture~\cite{Bedroya:2019snp, Bedroya:2020rmd} (see also~\cite{Brahma:2019vpl, Seo:2019wsh, Cai:2019dzj}).

\begin{figure}
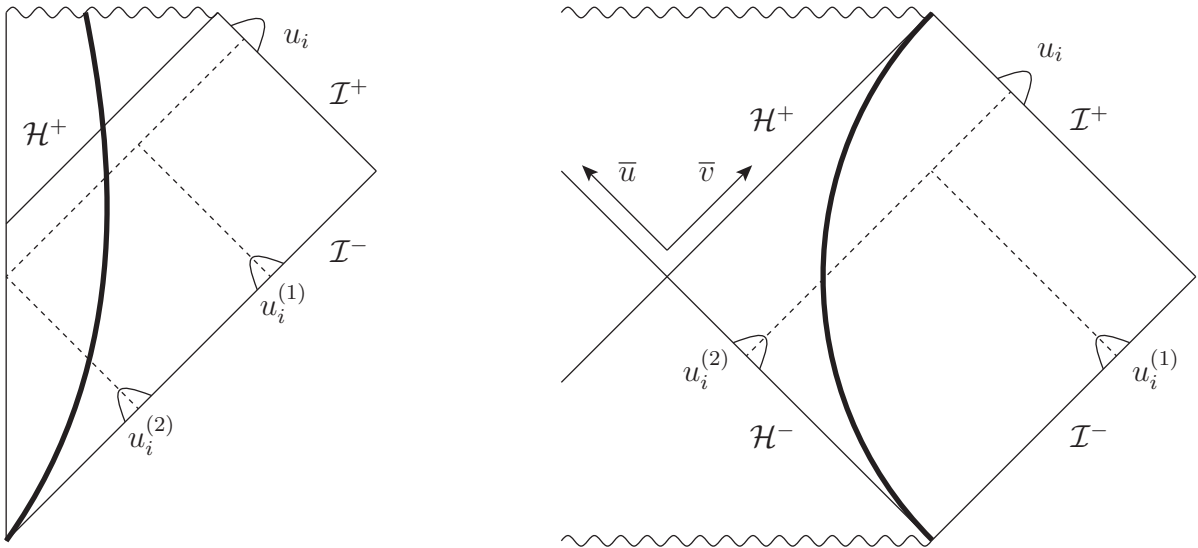

\begin{center}
 \begin{axopicture}(450,200)
  \Line(0,0)(0,200)
  \Line(0,0)(140,140)
  \Line(80,200)(140,140)
  \Line(80,200)(0,120)
  \Photon(0,200)(80,200){2}{8}
  \Bezier[width=2](00,0)(50,70)(40,150)(30,200)
  \DashLine(0,100)(90,190){2}
  \DashLine(0,100)(50,50){2}
  \DashLine(50,150)(100,100){2}
  \Photon(45,45)(55,55){10}{0.5}
  \Photon(95,95)(105,105){10}{0.5}
  \Photon(85,195)(95,185){10}{0.5}
  \Text(130,110)[]{$\mathcal{I}^-$}
  \Text(130,170)[]{$\mathcal{I}^+$}
  \Text(15,155)[]{$\mathcal{H}^+$}
  \Text(55,40)[]{$u_i^{(2)}$}
  \Text(105,90)[]{$u_i^{(1)}$}
  \Text(110,190)[]{$u_i$}
  \Line(450,100)(350,0)
  \Line(450,100)(350,200)
  \Line(350,0)(210,140)
  \Line(350,200)(210,60)
  \Photon(210,0)(350,0){2}{14}
  \Photon(210,200)(350,200){2}{14}
  \LongArrow(250,110)(220,140)
  \LongArrow(250,110)(280,140)
  \Arc[width=2](450,100)(141,135,225)
  \DashLine(280,70)(380,170){2}
  \DashLine(350,140)(420,70){2}
  \Photon(275,75)(285,65){10}{0.5}
  \Photon(415,65)(425,75){10}{0.5}
  \Photon(375,175)(385,165){10}{0.5}
  \Text(290,40)[]{$\mathcal{H}^-$}
  \Text(410,40)[]{$\mathcal{I}^-$}
  \Text(410,160)[]{$\mathcal{I}^+$}
  \Text(290,160)[]{$\mathcal{H}^+$}
  \Text(235,140)[]{$\overline{u}$}
  \Text(265,140)[]{$\overline{v}$}
  \Text(265,65)[]{$u_i^{(2)}$}
  \Text(435,65)[]{$u_i^{(1)}$}
  \Text(395,185)[]{$u_i$}
 \end{axopicture}
\end{center}
\caption{Diagram for a black hole evaporation (left panel) in the real collapsing star and (right panel) in the Schwarzschild space-time.}
\label{fig:BHevap}
\end{figure}

It should be noted here that the existence of an observer detecting the thermal radiation alone does {\it not} ensure  the instability of the background geometry. This can be seen in the Unruh effect, in which Minkowski space-time is stable even though an accelerating observer does detect thermal radiation~\cite{Fulling:1972md, Davies:1975, Unruh:1976db}. The thermal radiation here is an observer-dependent phenomenum thus cannot be connected to the energy flow resulting in the deformation of Minkowski space-time. This implies that to see if dS is deformed by the backreaction induced by thermal radiation, some other physical quantity that breaks the dS isometry needs to be introduced in addition. Such a quantity must be observer-independent, which in turn means that any observer must find it. To see this in detail, we recall the evaporation mechanism of a collapsing star, which corresponds to the real world black hole~\cite{Hawking:1974sw} as illustrated in the left panel of Figure~\ref{fig:BHevap}. The thermodynamic nature of a black hole is attributed to the fact that the Hilbert space of the ``in-states" defined at past infinity ${\cal I^-}$ and that of the ``out-states" defined at future infinity ${\cal I^+}$ admit different vacuum states, i.e., each of them is annihilated by the operator associated with different positive frequency mode function. More precisely, it is an effect of the evolution from $u_i^{(2)}$, a mixture of the positive and negative frequency modes at ${\cal I}^-$, which has travelled through the interior of the star described by the time-dependent geometry to $u_i$, the positive frequency mode at ${\cal I}^+$. Meanwhile, a mere reflection of the wavepacket outside the star (described by the static Schwarzschild geometry) from $u_i^{(1)}$ to $u_i$ is irrelevant to thermal behavior. In terms of (maximally extended) Schwarzschild space-time shown in the right panel of Figure~\ref{fig:BHevap}, such a different behavior between $u_i^{(1)}$ and $u_i^{(2)}$ can be easily understood by putting $u_i^{(1)}$ and $u_i^{(2)}$ on ${\cal I^-}$ and on the past horizon ${\cal H^-}$ respectively at which different positive modes that annihilate the vacuum are taken.

The vacuum satisfying the above requirements is called the Unruh vacuum, one of three possible vacua for the quantum fluctuations in the Schwarzschild space-time which have different thermal properties~\cite{Candelas:1980zt}:

\begin{itemize}

\item Boulware vacuum~\cite{Boulware:1974dm}: The vacuum annihilated by the positive modes with respect to the Schwarzschild time $t$. Since any states propagating from ${\cal I}^-$ and ${\cal H}^-$ are constructed from the same vacuum state as that at ${\cal I}^+$, an observer on ${\cal I}^+$ does not detect any thermal flux. On the other hand, the energy-momentum tensor diverges around the horizon thus regarded as an unphysical state\footnote{
Since the vacuum expectation values of the energy-momentum tensor components are typically divergent~\cite{Akhmedov:2020qxd}, they need to be renormalized by subtracting their values at the Boulware (Hartle-Hawking) vacuum at infinity (around the horizon). 
}~\cite{Christensen:1977jc}.

\item Hartle-Hawking vacuum~\cite{Hartle:1976tp}: This vacuum is annihilated by the positive modes with respect to $\overline{u}$ for the propagation from ${\cal H}^-$ and $\overline{v}$ for the propagation from ${\cal I}^-$\footnote{
Here, we define the tortoise coordinate by 
\begin{equation*}
dr_* = \frac{dr}{1-2GM/r} \, .
\end{equation*}
From this, the Kruskal-Szekeres null coordinates are defined by 
\begin{equation*}
\overline{u} \equiv -\exp \bigg( - \frac{t-r_*}{4M} \bigg) 
\quad \text{and} \quad
\overline{v} \equiv \exp \bigg( \frac{t+r_*}{4M} \bigg)
\, .
\end{equation*}
Then $\overline{u}$ and $\overline{v}$ correspond to the canonical affine parameters at ${\cal H}^-$ and ${\cal I}^-$, respectively. The Kruskal-Szekeres coordinates in dS will be defined in Section~\ref{sec:Bogol}.
}. 
This allows thermal flux on ${\cal I}^+$ propagated from both ${\cal I}^-$ and ${\cal H}^-$, but their contributions are cancelled with each other hence the net thermal flux vanishes. It describes thermal equilibrium between a black hole and thermal radiation.

\item Unruh vacuum~\cite{Unruh:1976db}: It is defined by taking the propagation from ${\cal I}^-$ to be positive frequency with respect to $t$ while that from ${\cal H}^-$ to be positive frequency with respect to $\overline{u}$ to annihilate the vacuum. As an observer on ${\cal I}^+$ detects the thermal flux from ${\cal H}^-$ but does not detect that from ${\cal I}^-$,  the net thermal flux does not vanish, which signals the dissipation of the energy of a black hole. Thus, black hole evaporation is well described in the Unruh vacuum.

\end{itemize}

Investigation of the black hole evaporation so far thus implies that the deformation of dS by thermal radiation is realized by a specific vacuum choice: In the Unruh vacuum, different positive modes that annihilate the vacuum are assigned to two independent propagations  such that the net thermal flux does not vanish. Of course, it is not straightforward to apply the results in black hole physics to dS, as the geometric structures of two space-times are different. A black hole is a compact object, so the asymptotic flat region ${\cal I}^\pm$ far from the event horizon is well defined globally and the energy flux passing through ${\cal I}^+$ is evidently interpreted as a decrease in the black hole mass. Moreover, the event horizon as a boundary of the black hole is an observer-independent object. In contrast, in dS the event horizon as a source of thermal radiation is an observer-dependent object surrounding the observer and the asymptotic flat region does not exist~\cite{Gibbons:1977mu, Witten:2001kn, Davies:2003me} [see, however, the discussion below \eqref{eq:Hevolution}]. Then a point on the horizon for some observer does not necessarily lie on the horizon for another observer and the vacuum conditions imposed on ${\cal H}^{\pm}$ are also observer-dependent statements. Nevertheless, we would like to stress that the vacuum conditions on the horizon is meaningful in the following sense. While the description of thermal radiation is based on the low-energy effective quantum field theory in the classical background, the dS isometry may be broken by some dynamics of which is not contained in the Wilsonian effective action\footnote{
It is argued in~\cite{Goheer:2002vf} that the dS isometry is not compatible with finite entropy. Since our computations consider the quadratic action only, such an dS isometry breaking effect is encoded in the defects in the background, which are generated by the coherent motion of the gravitons in a non-perturbative way. As an origin of the dS isometry breaking defects, we may consider a (self-)interacting scalar field~\cite{Polyakov:2012uc}. There, it is argued that the interaction gives rise to the violent particle production, which breaks the invariance of correlation functions under the dS isometry. This may be interpreted as a deformation of geometry through the strong backreaction. See, however, \cite{Starobinsky:1994bd, Marolf:2010nz, Hollands:2010pr} which claim that such violent particle production is compensated by the rapid expansion so that dS remains stable.
}.
This can be modeled by the scattering process in which the interaction between the radiated excitations and the ultraviolet dynamics takes place at some point inside the horizon, located far from both the observer and the horizon. Then as a result of such scattering process, \textit{any observer} will find that the background geometry is deformed. This is quantified by the change in the horizon area by investigating how the backreaction coming from a non-vanishing thermal flux affects the geometry through the semi-classically corrected Einstein equation\footnote{
Even though $\langle T_{\mu\nu}\rangle$ in the right-hand side is originated from quantum effects, it is classical in nature for it not to be averaged out. This is achieved by a process so called ``decoherence''~\cite{Zurek:2003zz, Schlosshauer:2003zy} by which we mean the non-unitary evolution due to the interaction between the system of our interest and the environment, causing the density matrix evolving from the pure state \cite{Paz:1991nd} (e.g. Bunch-Davies vacuum state) into the mixed state (see, however,~\cite{Polarski:1995jg}). For discussions on decoherence in (quasi-)dS, see e.g.,~\cite{Nelson:2016kjm, Gong:2019yyz}. We also note that decoherence does not mean that quantum nature disappears completely~\cite{Martin:2019wta, Gong:2020gdb}.
\label{footnote:deco}  
}
\begin{equation}
\label{eq:einstein}
G_{\mu\nu}+\Lambda g_{\mu\nu} = \frac{1}{\mpl^2} \langle T_{\mu\nu} \rangle
\, ,
\end{equation}
where $8\pi G = \mpl^{-2}$. That means even though the region that an observer can access is limited to the inside of the horizon, any observer can measure a non-vanishing thermal flux throughout the horizon, which is an \textit{observer-independent} phenomenon. But since the Wilsonian effective action is given only, an observer may model the origin of a non-vanishing thermal flux by imposing different vacuum conditions on ${\cal H}^-$ and ${\cal H}^+$. At least regarding our own observable Universe, this observation is relevant as the primordial inflation -- usually modeled by dS -- should have come to an end to commence the hot big bang evolution even if the physical origin of how to make a transition from dS to quasi-dS supported by the slow-roll of the inflaton is unclear. Based on this, the Unruh vacuum in two-dimensional dS was specified in~\cite{Aalsma:2019rpt} by assigning differently the positive modes at ${\cal H}^-$ and ${\cal H}^+$. Since two propagations move independently, they may have different positive modes that annihilate the vacuum, which can be used to realize the Unruh vacuum in a consistent way.

In this article, we study various vacua with different thermal properties in the four-dimensional dS with a scalar field, and investigate the stability of dS in the presence of thermal radiation from the quantum fluctuations of the scalar field. In Section~\ref{sec:modedS}, we argue that the mode function for the scalar field in four-dimensional dS can be separated into two independent modes by considering their behavior at the horizon. From this we calculate in Section~\ref{sec:Tmunu} the expectation values of the energy-momentum tensor components with respect to the Unruh vacuum, as well as the Boulware and the Hartle-Hawking vacua. Some details of calculation together with a brief summary on the Bogoliubov transformation can be found in Appendix~\ref{app:detsec3}. We also compare our results with those in other coordinate systems, the tortoise and the flat coordinates. The former shows how thermal flux is absorbed into or emitted from the horizon intuitively by defining the luminosity, and the latter is directly relevant to cosmology. In Section~\ref{sec:area} we consider the expansion $\Theta$ of the horizon to relate the non-vanishing thermal flux in the Unruh vacuum to the change in the horizon area through the backreaction. Several geometric properties used in this section are summarized in Appendix~\ref{app:Killhor}. Then we conclude in Section~\ref{sec:conclusion} with discussion on several issues on dS instability relevant to our results.

\section{Separation of modes in dS}
\label{sec:modedS}

In this section, we study in the space-time with horizon the mode function solution of a scalar field. We aim to separate the obtained solution into two different components, one being incoming from the horizon to the observer at the origin and the other being outgoing from her to the horizon. This enables us to associate different pairs of annihilation and creation operators with the incoming and outgoing modes, which is due to some dS isometrty breaking scattering process around dS inside the horizon. This in turn describes the instability of dS, as we will see in the following sections.

\subsection{Mode functions and vacua in Schwarzschild space-time}

\begin{figure}
\begin{center}
 \includegraphics[width=0.5\textwidth]{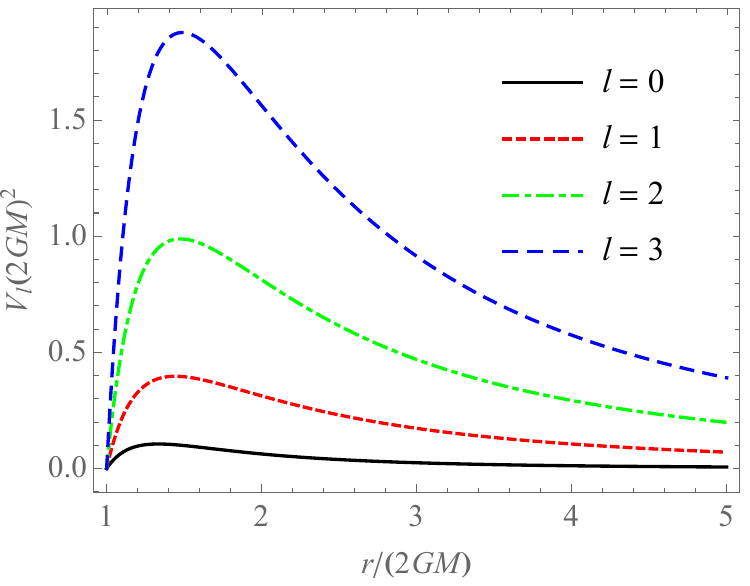}
\end{center}
\caption{The effective potential $V_\ell(r)$ for a massless scalar field in the Schwarzschild space-time. The maximum of $V_\ell(r)$ is at $r/(2GM) = 4/3$ ($\ell=0$), 1.443 ($\ell=1$), 1.476 ($\ell=2$), $\cdots$, 3/2 ($\ell\to\infty$).}
\label{fig:BHpot}
\end{figure}

\begin{figure}
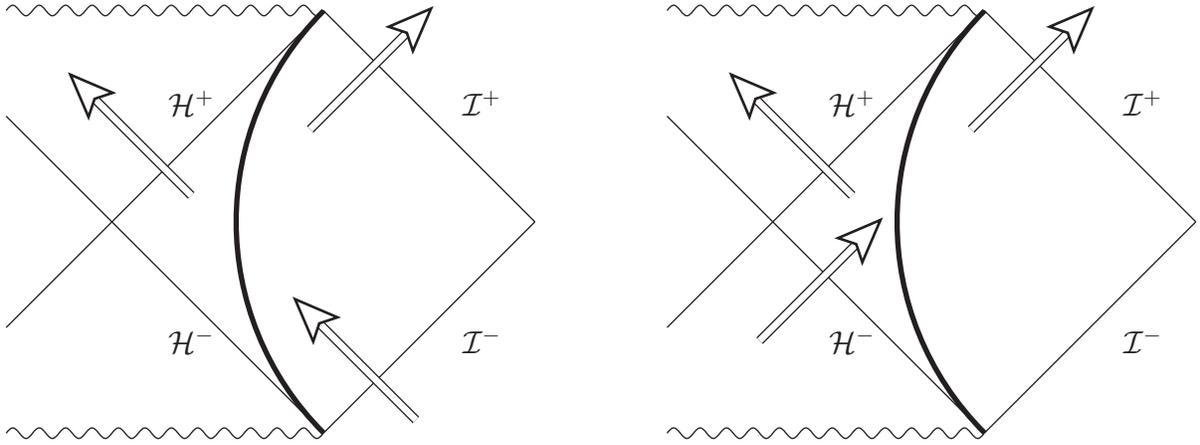

\begin{center}
 \begin{axopicture}(450,160)
  \Line(0,40)(120,160)
  \Line(0,120)(120,0)
  \Line(120,0)(200,80)
  \Line(120,160)(200,80)
  \Photon(0,160)(120,160){2}{12}
  \Photon(0,0)(120,0){2}{12}
  \Arc[width=2](200,80)(113,135,225)
  \Line[arrow,arrowpos=2,double,sep=3,arrowscale=1.5,arrowstroke=1](155,5)(115,45)
  \Line[arrow,arrowpos=2,double,sep=3,arrowscale=1.5,arrowstroke=1](70,90)(30,130)
  \Line[arrow,arrowpos=2,double,sep=3,arrowscale=1.5,arrowstroke=1](115,115)(155,155)
  \Text(180,35)[]{$\mathcal{I}^-$}
  \Text(180,125)[]{$\mathcal{I}^+$}
  \Text(70,35)[]{$\mathcal{H}^-$}
  \Text(70,125)[]{$\mathcal{H}^+$}
  \Line(250,40)(370,160)
  \Line(250,120)(370,0)
  \Line(370,0)(450,80)
  \Line(370,160)(450,80)
  \Photon(250,160)(370,160){2}{12}
  \Photon(250,0)(370,0){2}{12}
  \Arc[width=2](450,80)(113,135,225)
  \Line[arrow,arrowpos=2,double,sep=3,arrowscale=1.5,arrowstroke=1](320,90)(280,130)
  \Line[arrow,arrowpos=2,double,sep=3,arrowscale=1.5,arrowstroke=1](285,35)(325,75)
  \Line[arrow,arrowpos=2,double,sep=3,arrowscale=1.5,arrowstroke=1](365,115)(405,155)
  \Text(430,35)[]{$\mathcal{I}^-$}
  \Text(430,125)[]{$\mathcal{I}^+$}
  \Text(320,35)[]{$\mathcal{H}^-$}
  \Text(320,125)[]{$\mathcal{H}^+$}
 \end{axopicture}
\end{center}
\caption{Bahavior of the (left panel) ``in"- and the (right panel) ``up"-mode in the Schwarzschild space-time. The thick line indicates the location at which the effective potential is maximized, $r\approx3GM$.}
\label{fig:BHwave}
\end{figure}

Before discussing mode functions and vacua in dS, it is instructive to recall how these issues are visited in the Schwarzschild space-time. As will be clear shortly, the crucial difference between dS and Schwarzschild space-time arises from whether the asymptotic flat region exists. The Schwarzschild metric which describes the exterior of a black hole is given by
\begin{equation}
ds^2 = - \bigg( 1 - \frac{2GM}{r} \bigg) dt^2 + \frac{1}{1 - 2GM/r} dr^2 
+ r^2 \big( d\theta^2 + \sin^2\theta d\phi^2 \big)
\, .
\end{equation}
A minimally coupled massive scalar field with mass $m$ then satisfies the Klein-Gordon equation $(-\square +m^2)\varphi=0$ with $\square \equiv g^{\mu\nu}\partial_\mu\partial_\nu$. The solution corresponding to the positive frequency mode with respect to the Schwarzschild time $t$ can be written in the form of $\varphi=R_{\omega\ell}(r)Y_\ell^m(\theta, \phi)e^{-i\omega t}$, then $\chi_{\omega\ell}\equiv r R_{\omega\ell}$ satisfies the Schr\"odinger-like equation:
\begin{equation}
\label{eq:Schrodinger}
-\frac{d^2}{dr_*^2} \chi_{\omega\ell} + V_\ell (r) \chi_{\omega\ell} = \omega^2 \chi_{\omega\ell} \, ,
\end{equation}
where the effective potential $V_\ell(r)$ is given by
\begin{equation}
V_\ell (r) = \bigg( 1 - \frac{2GM}{r} \bigg) \bigg[ \frac{\ell(\ell+1)}{r^2} + m^2 + \frac{2GM}{r^3} \bigg] \, .
\end{equation}
The effective potential $V_\ell (r)$ has a maximum at $r \approx 3GM$ as depicted in Figure~\ref{fig:BHpot}, which divides the region outside the event horizon into the exterior of a black hole ($r \gtrsim 3GM$) and the thermal atmosphere ($2GM<r \lesssim 3GM$) \cite{Zurek:1985gd, Thorne:1986iy} (for a recent discussion, see e.g.,~\cite{Nomura:2018kia}). From this we expect that $R_{\omega\ell}$ behaves like the wave in the scattering process: Initially the wave is injected from either $r=\infty$ (${\cal I}^-$) or the horizon $r=2GM$ (${\cal H}^-$), then is reflected or transmitted by the potential barrier produced by the background geometry. Different initial conditions enable one to take the ``in"-mode $R_{\omega\ell}^{\rm in}$ (${\cal I}^- \to {\cal H}^+/{\cal I}^+$) and the ``up"-mode $R_{\omega\ell}^{\rm up}$ (${\cal H}^- \to {\cal H}^+/{\cal I}^+$) to form the independent basis set of the positive mode functions as shown in Figure~\ref{fig:BHwave}~\cite{DeWitt:1975ys}.

For the field quantization, we define a pair of the annihilation and creation operators $(b_{\rm in}, b_{\rm in}^\dagger)$ which supports $R_{\omega\ell}^{\rm in}$ and $(b_{\rm up}, b_{\rm up}^\dagger)$ which supports $R_{\omega\ell}^{\rm up}$ as the positive frequency modes  with respect to $t$. Then $b_{\rm in}$ and $b_{\rm up}$ annihilate the Boulware vacuum. On the other hand, we can also define the annihilation and creation operators $(a_{\overline{u}}, a_{\overline{u}}^\dagger)$ and $(a_{\overline{v}}, a_{\overline{v}}^\dagger)$ associated with the positive frequency modes with respect to the Kruskal-Szekeres null coordinates $\overline{u}$ and $\overline{v}$. The Hartle-Hawking vacuum is annihilated by $a_{\overline{u}}$ for the up-mode and $a_{\overline{v}}$ for the in-mode. Since $R_{\omega\ell}^{\rm in}$ and $R_{\omega\ell}^{\rm up}$ are the positive frequency modes with respect to $t$ but the mixtures of the positive and the negative frequency modes with respect to $\overline{u}$ and $\overline{v}$, an observer on ${\cal I}^+$ finds that both the in- and up-modes have thermal character. Finally, the Unruh vacuum is annihilated by $(b_{\rm in}, b_{\rm in}^\dagger)$ for the in-mode but by $(a_{\overline{u}}, a_{\overline{u}}^\dagger)$ for the up-mode, such that an observer on ${\cal I}^+$ finds thermal character in the up-mode only~\cite{Candelas:1980zt}.

\subsection{Separation of modes in the dS static coordinates}
\label{subsec:dSsep}

\begin{figure}
\begin{center}
 \includegraphics[width=0.5\textwidth]{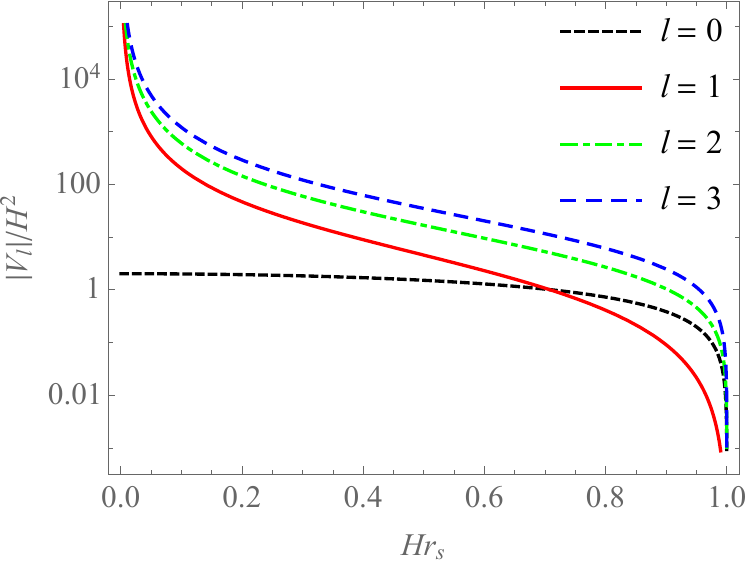}
\end{center}
\caption{The effective potential $V_\ell(r_s)$ for a massless ($m=0$), minimally coupled ($\xi=0$) scalar field in dS. Except for $\ell=0$ for which $V_{\ell=0}(r_s) < 0$, $V_\ell(r_s)$ diverges as $r_s\to0$.}
\label{fig:dSwave}
\end{figure}

We now consider dS in the static coordinates. This is because by adopting the static coordinates the existence of the horizon, separated from an observer by $1/H$, becomes manifest in the metric so that studying thermal effects from the horizon is sensible. The dS metric in terms of the static coordinates is given by
\begin{equation}
\label{eq:dSstatic}
ds^2 = - \big( 1-H^2 r_s^2 \big) dt_s^2 + \frac{dr_s^2}{1-H^2r_s^2} 
+ r_s^2 \big( d\theta^2 + \sin^2\theta d\phi^2 \big) 
\, .
\end{equation}
A scalar field $\varphi$ in dS satisfies the Klein-Gordon equation:
\begin{equation}
\label{eq:wave}
\bigg\{ \frac{1}{1-H^2 r_s^2} \partial_{t_s}^2 
- \frac{1}{r_s^2} \partial_{r_s} \Big[ r_s^2 \big( 1 - H^2r_s^2 \big) \partial_{r_s} \Big] 
- \frac{1}{r_s^2} \nabla^2_{\Omega} + m^2 + \xi \calR \bigg\} \varphi
= 0 \, ,
\end{equation}
where $\nabla^2_{\Omega}$ is the Laplacian for the angular variables, and $\xi$ is the non-minimal coupling to Ricci scalar ${\cal R}=12H^2$ ($\xi=1/6$ for a conformally coupled scalar field). Just like in the Schwarzschild space-time, the solution corresponding to the positive frequency mode with respect to $t_s$ is written as $\varphi(t_s, r_s, \theta, \phi)=C_{\omega\ell} R_{\omega  \ell}(H r_s)Y_{\ell}^m(\theta, \phi)e^{-i\omega t_s}$. Then $\chi_{\omega\ell}\equiv r_s R_{\omega\ell}$ satisfies the Schr\"odinger-like equation \eqref{eq:Schrodinger}, with the effective potential given by
\begin{equation}
\label{eq:effpot}
V_\ell (r_s) = \big( 1 - H^2 r_s^2 \big) \bigg[ \frac{\ell(\ell+1)}{r_s^2} 
+ \big( m^2 + 12 \xi H^2 \big) - 2 H^2 \bigg] \, .
\end{equation}
For $\ell\neq0$, the potential becomes infinitely high around $r_s=0$ as shown in Figure~\ref{fig:dSwave}. Comparing this with the effective potential in the Schwarzschild space-time, we realize that dS has only thermal atmosphere (the region between the horizon and the maximum of the effective potential) but the region corresponding to the black hole exterior does not exist~\cite{Nomura:2019qps}. Thus, we expect that  $R_{\omega\ell}$ immediately decays around $r_s=0$. The potential wall at $r_s=0$ does not appear for $\ell=0$ but still, $\chi_{\omega\ell} \approx 0$ hence $R_{\omega\ell}$ does not oscillate but converges to a constant at $r_s=0$: This will soon be found explicitly from \eqref{eq:radial}, which gives $R_{\omega\ell} \approx (H r_s)^\ell$ as $r_s \to 0$. Meanwhile, $R_{\omega\ell}$ behaves like the superposition of the incoming and outgoing plane waves around the horizon, where $V_\ell(r_s)$ vanishes.

Solving \eqref{eq:wave}, the radial part of the normalizable solution (the solution which diverges at $r_s =0$ is excluded) is given by
\begin{align}
\label{eq:radial}
R_{\omega \ell}(x)
& =
x^\ell (1-x^2)^{i\omega/(2H)}
F \bigg( \frac{\ell+3/2+\nu+i\omega/H}{2}, \frac{\ell+3/2-\nu+i\omega/H}{2} ; \ell+\frac32 ; x^2 \bigg)
\nonumber\\
& =
y^\ell (1+y^2)^{(3+2\nu)/4} 
F \bigg( \frac{\ell+3/2+\nu+i\omega/H}{2}, \frac{\ell+3/2+\nu-i\omega/H}{2} ; \ell+\frac32 ; -y^2 \bigg)
\, ,
\end{align}
where
\begin{equation}
x \equiv Hr_s \, ,
\qquad
y \equiv \frac{x}{\sqrt{1-x^2}} \, ,
\qquad
\nu \equiv \frac32 \sqrt{ 1 - \frac49 \bigg( \frac{m^2}{H^2} + 12\xi \bigg)} \, .
\end{equation}
We note here the linear transformation formula for the hypergeometric function (e.g., 15.3.7 of~\cite{Abramowitz}),
\begin{align}
\label{eq:hyper}
F(a,b;c;-z)
& =
\frac{\Gamma(c)\Gamma(b-a)}{\Gamma(b)\Gamma(c-a)} z^{-a} F\bigg(a,1-c+a;1-b+a;-\frac{1}{z}\bigg)
\nonumber\\
& \quad
+ \frac{\Gamma(c)\Gamma(a-b)}{\Gamma(a)\Gamma(c-b)} z^{-b} F\bigg(b,1-c+b;1-a+b;-\frac{1}{z}\bigg) 
\, ,
\end{align}
then in the $z\to \infty$ limit 
\begin{equation}
\label{eq:hyperapp}
F(a,b;c;-z) \underset{z\to\infty}{\longrightarrow} 
\Gamma(c) \bigg[ \frac{\Gamma(b-a)}{\Gamma(b)\Gamma(c-a)} z^{-a}
+ \frac{\Gamma(a-b)}{\Gamma(a)\Gamma(c-b)} z^{-b} \bigg]
\, ,
\end{equation}
since $F(a,b;c;w)\to 1$ as $w\to 0$. Replacing the hypergeometric function in the second line of \eqref{eq:radial} by the separated form \eqref{eq:hyper}, we can write $R_{\omega\ell}(x)$ as a sum of two functions of $r_s$, namely, the incoming and outgoing modes:
\begin{equation}
R_{\omega\ell}(x) = R^{\rm in}_{\omega\ell}(x) + R^{\rm out}_{\omega\ell}(x) \, .
\end{equation}
This separation of the mode function has a clear interpretation around the horizon. To see this, we take the limit $x\to 1$ or equivalently $y \to \infty$, which gives
\begin{equation}
\label{eq:horwave}
\begin{split}
R^{\rm in}_{\omega\ell}(x) 
& \underset{x\to 1}{\longrightarrow} 
A_{\omega \ell}^*(1-x^2)^{i\omega/(2H)} 
\simeq 
A_{\omega\ell}^* e^{i\omega\log2/H}e^{-i\omega r_*}
\, ,
\\
R^{\rm out}_{\omega\ell}(x)
& \underset{x\to 1}{\longrightarrow} 
A_{\omega\ell}(1-x^2)^{-i\omega/(2H)} 
\simeq 
A_{\omega\ell} e^{-i\omega\log2/H}e^{i\omega r_*}
\, ,
\end{split}
\end{equation}
where the coefficient $A_{\omega\ell}$ is given by
\begin{equation}
A_{\omega\ell}
\equiv
\frac{\Gamma(\ell+3/2) \Gamma(i\omega/H)}
{\Gamma\Big(\frac{\ell+3/2+\nu+i\omega/H}{2}\Big) \Gamma\Big(\frac{\ell+3/2-\nu+i\omega/H}{2}\Big) } 
\, ,
\end{equation}
and we have used in the last step the tortoise coordinate:
\begin{equation}
dr_* = \frac{dr_s}{1 - H^2 r_s^2}
\quad \text{or} \quad
r_*=\frac{1}{2H} \log\Big(\frac{1+ H r_s}{1-H r_s}\Big) \, .
\end{equation}
Therefore, combined with the time dependence $\exp(-i\omega t_s)$, $R^{\rm in}_{\omega\ell}$ and $R^{\rm out}_{\omega\ell}$ describe respectively the incoming and outgoing propagations around the horizon.

Indeed, the hypergeometric differential equation has three pairs of fundamental solutions associated with three regular singularities, which are extended to the Kummer's 24 solutions~\cite{Abramowitz, Whittaker}. The two terms in \eqref{eq:hyper} in fact form one of the fundamental solution pairs, thus they solve the hypergeometric differential equation independently. This implies that $R^{\rm in}_{\omega\ell}$ and $R^{\rm out}_{\omega\ell}$ can be treated as two {\it independent} solutions to the wave equation \eqref{eq:wave}, as the in- and up-modes form a complete, independent basis set of solutions to the wave equation in the Schwarzschild space-time. Meanwhile, it is important to note that, unlike $R_{\omega\ell} = R^{\rm in}_{\omega\ell}+R^{\rm out}_{\omega\ell}$ which is regular {\it everywhere} inside the horizon, both $R^{\rm in}_{\omega\ell}$ and $R^{\rm out}_{\omega\ell}$ are not well-defined but diverge at $r_s=0$ and thus are not normalizable. Hence it is not sensible to take $R^{\rm in}_{\omega\ell}$ and $R^{\rm out}_{\omega\ell}$ as two independent modes over the {\it whole} region inside the horizon. However, as we have argued in the introduction, the interaction between the dS isometry breaking effects and the mode excitations can be modeled by the scattering process that takes place at some point inside the horizon, mimicking the scattering around the potential wall in the black hole. Thus, imposing the asymptotic dS geometry at $r_s=0$ and $r_s=H^{-1}$ at leading order, the incoming and outgoing wavefunctions, which behave due to this ``scattering'' as
\begin{equation}
\label{eq:horwave}
\psi^{\rm in (out)}_{\omega\ell}(x) 
\longrightarrow 
\left\{
\begin{array}{ll}
R^{\rm in (out)}_{\omega\ell}(x) & \text{as} \quad x\to 1 \quad (r_s \to H^{-1})
\vspace{0.5em}
\\
R^{\rm in}_{\omega\ell}(x)+R^{\rm out}_{\omega\ell}(x) & \text{as} \quad x\to 0 \quad (r_s \to 0)
\end{array}
\right.
\, ,
\end{equation}
become normalizable solutions and are independent around the horizon. As can be seen in Appendix \ref{app:normalization}, since the integral for the normalization of the wavefunction is mainly contributed from $r_s\simeq H^{-1}$, we can take $\psi^{\rm in}_{\omega \ell}$ and $\psi^{\rm out}_{\omega \ell}$ as two (almost) independent modes. Thus when we focus on the horizon, instead of assigning a single pair of annihilation and creation operators $(b_{\omega \ell m}, b_{\omega \ell m}^\dagger)$ to $R_{\omega\ell}$ as a whole, we can associate two {\it independent} pairs of annihilation and creation operators to $R^{\rm in}_{\omega\ell}$ and $R^{\rm out}_{\omega\ell}$ separately:
\begin{equation}
\big( b_{\omega \ell m}^{\rm in}, b_{\omega \ell m}^{{\rm in}, \dagger} \big) 
\leftrightarrow R^{\rm in}_{\omega\ell}
\quad \text{and} \quad
\big( b_{\omega \ell m}^{\rm out}, b_{\omega \ell m}^{{\rm out}, \dagger} \big) \leftrightarrow 
R^{\rm out}_{\omega\ell}
\, .
\end{equation}

We close this section by noting that even though the dS metric in the static coordinates is reduced to the Minkowski one at $r_s=0$, what an observer at $r_s=0$ sees is quite different from what an observer on ${\cal I}^+$ of the Schwarzschild space-time sees. Indeed, the flatness of the metric \eqref{eq:dSstatic} at $r_s=0$ is just an artifact of the coordinate choice, which is evident from the fact that the dS Ricci scalar ${\cal R}=12 H^2$ is constant everywhere. On the contrary, the asymptotic flatness in the Schwarzschild space-time is physical since the curvature tensor vanishes as $r\to\infty$. This is reflected in the shape of the dS effective potential \eqref{eq:effpot}, which does not vanish but diverges at $r_s=0$. That is, the effective potential only describes thermal atmosphere, but the region corresponding to the exterior of a black hole which contains the asymptotic flat infinity is not defined. As a result, $R_{\omega\ell}$ does not propagate but decays (for $\ell \ne 0$) or converges to a constant value (for $\ell = 0$) at $r_s=0$, and the propagation of the plane wave appears around the horizon only. This is the reason why the thermal properties of dS in this article will be discussed in terms of the modes propagating into ${\cal H}^+$ ($R^{\rm in}_{\omega\ell}$) or from ${\cal H}^-$ ($R^{\rm out}_{\omega\ell}$), instead of considering the modes propagating from or to an observer at $r_s=0$. Both $R^{\rm in}_{\omega\ell}$ and $R^{\rm out}_{\omega\ell}$ are the positive frequency modes with respect to the static time coordinate $t_s$, the proper time of an observer at $r_s=0$, hence the mixture of the positive and negative frequency modes with respect to the Kruskal-Szekeres null coordinates $\overline{u}$ or $\overline{v}$. This leads to the thermodynamic description around the horizon.

\begin{figure}
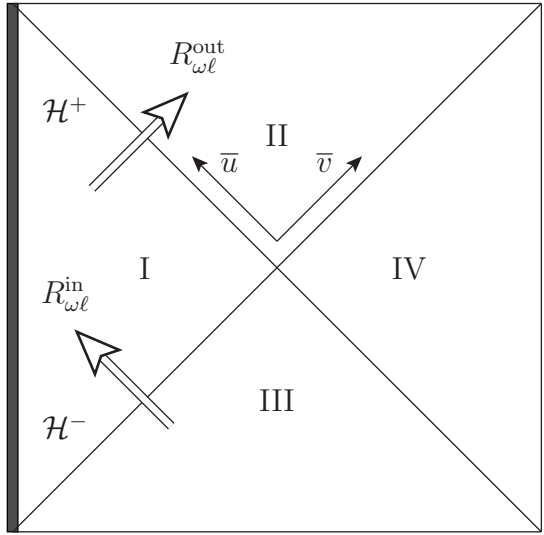

\begin{center}
 \begin{axopicture}(200,200)
  \GBox(-2,0)(2,200){0.3}
  \Line(0,0)(200,0)
  \Line(200,0)(200,200)
  \Line(0,200)(200,200)
  \Line(0,0)(200,200)
  \Line(0,200)(200,0)
  \LongArrow(100,110)(70,140)
  \LongArrow(100,110)(130,140)
  \Line[arrow,arrowpos=2,double,sep=3,arrowscale=1.5,arrowstroke=1](60,40)(30,70)
  \Line[arrow,arrowpos=2,double,sep=3,arrowscale=1.5,arrowstroke=1](30,130)(60,160)
  \Text(50,100)[]{I}
  \Text(150,100)[]{IV}
  \Text(100,50)[]{III}
  \Text(100,150)[]{II}
  \Text(82,140)[]{$\overline{u}$}
  \Text(118,140)[]{$\overline{v}$}
  \Text(20,40)[]{$\mathcal{H}^-$}
  \Text(20,160)[]{$\mathcal{H}^+$}
  \Text(20,90)[]{$R_{\omega\ell}^\text{in}$}
  \Text(70,180)[]{$R_{\omega\ell}^\text{out}$}
 \end{axopicture}
\end{center}
\caption{Penrose diagram of dS. An observer at rest is located at $r_s=0$ of Region I (leftmost side), in which $\overline{u}>0$ and $\overline{v}<0$. We also show the behavior of the plane wave solutions in dS. The thick vertical line indicates the effective potential wall around $r_s=0$. }
\label{fig:dSPenrose}
\end{figure}

\section{Energy-momentum tensor in dS vacua}
\label{sec:Tmunu}

In the previous section, we have seen that the mode function in dS can be separated into the incoming and outgoing modes, with different pairs of annihilation and creation operators. This models the effects of   physics responsible for the instability of dS as scattering process inside the horizon. In this section, we compute the corresponding energy-momentum tensor based on the separation of the modes discussed in the previous section. We then show that a non-vanishing transfer of thermal energy is possible only in the Unruh vacuum among different possible vacua. This non-zero ``flux'' of thermal energy through the horizon leads to the change in the dS geometry, as will be discussed in the following section.

\subsection{Bogoliubov transformation}
\label{sec:Bogol}

The static coordinates as defined in \eqref{eq:dSstatic} just cover Region I in Figure~\ref{fig:dSPenrose} which is surrounded by the past (${\cal H}^-$) and the future (${\cal H}^+$) horizon. On the other hand, the extension to the whole dS manifold can be achieved through the Kruskal-Szekeres null coordinates. In Region I, they are defined in terms of the Eddington-Finkelstein coordinates $u=t_s-r_*$ and $v=t_s+r_*$ by\footnote{
Our definition coincides with~\cite{Spradlin:2001pw}, which denotes propagations incoming ``toward" and outgoing ``from" $r_s=0$ in a consistent way. In~\cite{Gibbons:1977mu, Aalsma:2019rpt}, $\overline{u}$ ($\overline{v}$) is defined in terms of $t_s+r_*$ ($t_s-r_*$). These definitions are advantageous in specifying the directions of propagation with respect to the inaccessible region, i.e., whether the wave propagates into or out of the horizon.
}
\begin{equation}
\label{eq:statKS}
\begin{split}
\overline{u}
& \equiv
\frac{1}{H} e^{Hu}
= 
\frac{1}{H} e^{Ht_s} \sqrt{\frac{1-Hr_s}{1+Hr_s}}
\, ,
\\ 
\overline{v}
& \equiv
- \frac{1}{H} e^{-H v}
=
-\frac{1}{H} e^{-Ht_s} \sqrt{\frac{1-Hr_s}{1+Hr_s}}
\, ,
\end{split}
\end{equation}
which give
\begin{equation}
\label{eq:KSstat}
Hr_s = \frac{1+H^2\overline{u}\overline{v}}{1-H^2\overline{u}\overline{v}}
\quad \text{and} \quad
e^{2Ht_s} = -\frac{\overline{u}}{\overline{v}}
\, ,
\end{equation}
so that the dS metric \eqref{eq:dSstatic} is written as
\begin{equation}
\label{eq:dSkruskal}
ds^2 
=
- \frac{4}{(1-H^2\overline{u}\overline{v})^2} d\overline{u} d\overline{v}
+ \frac{(1+H^2\overline{u}\overline{v})^2}{H^2(1-H^2\overline{u}\overline{v})^2}
\big( d\theta^2 + \sin^2\theta d\phi^2 \big)
\, . 
\end{equation}
We note that the time-like Killing vector in dS is given by $k^a = (\partial_{t_s})^a = H \big( \overline{u}\partial_{\overline u} - \overline{v}\partial_{\overline v} \big)^a$. The future horizon ${\cal H}^+$ is a null hypersurface satisfying $\overline{v}=0$, on which $k^a=H\overline{u}(\partial_{\overline{u}})^a$ is normal to ${\cal H}^+$ as can be easily noticed from $k_a=-2H\overline{u}(d\overline{v})_a$. In the same way, the past horizon ${\cal H}^-$ is a null hypersurface $\overline{u}=0$, on which $k^a=-H\overline{v} (\partial_{\overline{v}})^a$, or $k_a=2H\overline{v}(d\overline{u})_a$, is normal to ${\cal H}^-$. This indicates that $\overline{u}$ ($\overline{v}$) becomes the canonical affine parameter on ${\cal H}^+$ (${\cal H}^-$).

The thermodynamic properties of the horizon is quantified by the Bogoliubov transformation (see Appendix~\ref{app:Bogoliubov} for a brief summary) between the positive frequency mode with respect to $t_s$ and that with respect to the Kruskal-Szekeres null coordinates~\cite{Crispino:2007eb}. To see this in detail, we list the extension of the Kruskal-Szekeres null coordinates and approximations   of $R^{\rm in}_{\omega\ell}$ and $R^{\rm out}_{\omega\ell}$ around the horizon below, which is justified in Appendix~\ref{app:normalization}. Hereafter the symbol $\simeq$ stands for taking the horizon limit ($r_s\to H^{-1}$) of the given function: 
\begin{align}
\label{eq:regI-asymptoticsol1}
\text{Region I}
: \quad & 
\overline{u} = \frac{1}{H} e^{H u} \quad \text{and} \quad \overline{v} = - \frac{1}{H} e^{-Hv}
\nonumber\\
&
R^{\rm out, I}_{\omega \ell} 
\simeq 
\theta(\overline{u}) \frac{e^{i\delta_{\omega\ell}}H}{\sqrt{4\pi\omega}} e^{-i\omega u}
=
\theta(\overline{u}) \frac{e^{i\delta_{\omega\ell}}H}{\sqrt{4\pi\omega}} (H\overline{u})^{-i\omega/H}
\, ,
\\
&
R^{\rm in, I}_{\omega \ell} 
\simeq 
\theta(-\overline{v}) \frac{e^{-i\delta_{\omega\ell}}H}{\sqrt{4\pi\omega}} e^{-i\omega v}
=
\theta(-\overline{v}) \frac{e^{-i\delta_{\omega\ell}}H}{\sqrt{4\pi\omega}} (-H\overline{v})^{i\omega/H}
\, ,
\nonumber\\
\label{eq:regI-asymptoticsol2}
\text{Region IV}
: \quad &
\overline{u} = - \frac{1}{H} e^{Hu} \quad \text{and} \quad \overline{v} = \frac{1}{H} e^{-Hv}
\nonumber\\
&
R^{\rm out, IV}_{\omega \ell} 
\simeq 
\theta(-\overline{u}) \frac{e^{-i\delta_{\omega\ell}}H}{\sqrt{4\pi\omega}} e^{i\omega u}
=
\theta(-\overline{u}) \frac{e^{-i\delta_{\omega\ell}}H}{\sqrt{4\pi\omega}} (-H\overline{u})^{i\omega/H}
\, ,
\\
& 
R^{\rm in, IV}_{\omega \ell} 
\simeq 
\theta(\overline{v}) \frac{e^{i\delta_{\omega\ell}}H}{\sqrt{4\pi\omega}} e^{i\omega v}
=
\theta(\overline{v}) \frac{e^{i\delta_{\omega\ell}}H}{\sqrt{4\pi\omega}} (H\overline{v})^{-i\omega/H}
\, ,
\nonumber
\end{align}
where $\delta_{\omega\ell} = \text{arg}[A_{\omega\ell}] - i\omega\log2/H$.
Here the positive mode in Region IV is taken to be $\exp(i\omega u)$, rather than $\exp(-i\omega u)$, because the time-like Killing vector in this region is $-\partial_{t_s}$.

We now focus on the Bogoliubov transformation of the outgoing mode, $R^{\rm out}_{\omega \ell}$. We first note that around the horizon, $\big( R^{\rm out, IV}_{\omega \ell} \big)^*$ is written as
\begin{equation}
\big( R^{\rm out, IV}_{\omega \ell} \big)^* 
\simeq 
\theta(-\overline{u}) \frac{e^{i\delta_{\omega\ell}}H}{\sqrt{4\pi\omega}} 
e^{-i\frac{\omega}{H} \log(-H\overline{u})}
=
\theta(-\overline{u}) \frac{e^{i\delta_{\omega\ell}}H}{\sqrt{4\pi\omega}} 
e^{\frac{\pi \omega}{H}} e^{-i\frac{\omega}{H} \log(H\overline{u})}
\, ,
\end{equation}
where the function of complex $\overline{u}$ is analytic in the lower half plane, such that the branch cut extends over the positive imaginary axis. As $|H\overline{u}|=-H\overline{u}$ in Region IV, we have $\log(H\overline{u}) = \log \big( |H\overline{u}| \big) - i\pi$. From this, we obtain the positive frequency mode functions with respect to $\overline{u}$ as
\begin{equation}
\label{eq:Bogol}
\begin{split}
R^{\rm out, (1)}_{\omega \ell}
& =
\frac{1}{\sqrt{2\sinh(\pi\omega/H)}} 
\Big[ e^{\frac{\pi\omega}{2H}} R^{\rm out, I}_{\omega \ell}  
+ e^{-\frac{\pi\omega}{2H}} \big( R^{\rm out, IV}_{\omega \ell} \big)^* \Big] 
\sim 
e^{-i\frac{\omega}{H}\log(H\overline{u})}
\, ,
\\
R^{\rm out, (2)}_{\omega \ell}
& =
\frac{1}{\sqrt{2\sinh(\pi\omega/H)}} 
\Big[ e^{-\frac{\pi\omega}{2H}} \big( R^{\rm out, I}_{\omega \ell} \big)^*  
+ e^{\frac{\pi\omega}{2H}} R^{\rm out, IV}_{\omega \ell} \Big] 
\sim 
e^{+i\frac{\omega}{H}\log(H\overline{u})}
\, ,
\end{split}
\end{equation}
which define the Bogoliubov transformation between $R^{\rm out, I/IV}_{\omega \ell}$, the positive frequency modes with respect to $t_s$, and $R^{\rm out, (1)/(2)}_{\omega \ell}$, those with respect to $\overline{u}$. Now we associate the annihilation and creation operators with the mode functions as
\begin{equation}
\Big( b_{\omega \ell m}^{\rm out I/IV}, b_{\omega \ell m}^{{\rm out I/IV},\dagger} \Big) 
\leftrightarrow R^{\rm out, I/IV}_{\omega \ell}
\quad \text{and} \quad
\Big( a_{\omega \ell m}^{{\rm out}(1)/(2)}, a_{\omega \ell m}^{{\rm out (1)/(2)},\dagger} \Big) 
\leftrightarrow R^{\rm out, (1)/(2)}_{\omega \ell}
\, .
\end{equation}
Then the inverse of the Bogoliubov transformation above gives
\begin{equation}
\label{eq:Bogolfin}
\begin{split}
R^{\rm out, I}_{\omega \ell}
& = 
\frac{1}{\sqrt{2\sinh(\pi\omega/H)}} 
\Big[ e^{\frac{\pi\omega}{2H}} R^{\rm out, (1)}_{\omega \ell} 
- e^{-\frac{\pi\omega}{2H}} \big( R^{\rm out, (2)}_{\omega \ell} \big)^* \Big]
\, ,
\\
b^{\rm out, I}_{\omega \ell m}
& =
\frac{1}{\sqrt{2\sinh(\pi\omega/H)}} 
\Big[ e^{\frac{\pi\omega}{2H}} a^{{\rm out}(1)}_{\omega \ell m}  
+ e^{-\frac{\pi\omega}{2H}} \big( a^{{\rm out}(2)}_{\omega \ell m} \big)^\dagger \Big] 
\, .
\end{split}
\end{equation}
By considering the incoming mode as a function of $\overline{v}$ in the same way, we also obtain the Bogoliubov transformation for $R^{\rm in, I}_{\omega \ell}$ and $b^{\rm in, I}_{\omega \ell m}$ as well. Since we are interested in Region I, we will omit ``I" representing Region I in the superscript of the annihilation and creation operators from now on.

With these two different sets of the annihilation and creation operators that support different  positive frequency modes, we can define different types of vacua analogous to those in the Schwarzschild space-time~\cite{Aalsma:2019rpt}:
\begin{itemize}

\item Boulware vacuum $|B\rangle$: The vacuum state annihilated by both $b^{\rm in}_{\omega \ell m}$ and $b^{\rm out}_{\omega \ell m}$.

\item Hartle-Hawking vacuum $|H\rangle$: The vacuum state annihilated by both $a^{\rm in}_{\omega \ell m}$ and $a^{\rm out}_{\omega \ell m}$. As will be clear in Section~\ref{sec:flatcoord}, this in fact corresponds to the Bunch-Davies vacuum~\cite{Chernikov:1968zm, Bunch:1978yq}, which preserves the dS isometry. 

\item Unruh vacuum $|U\rangle$: The vacuum state annihilated by $a^{\rm in}_{\omega \ell m}$ and $b^{\rm out}_{\omega \ell m}$. We may define another Unruh vacuum $|U'\rangle$ that is annihilated by $a^{\rm out}_{\omega \ell m}$ and $b^{\rm in}_{\omega \ell m}$.

\end{itemize}

\subsection{Energy-momentum tensor around the horizon}
\label{subsec:Tmunuhor}

In Section~\ref{subsec:dSsep}, we have argued that $R^{\rm in}_{\omega \ell}$ and $R^{\rm out}_{\omega \ell}$ in Region I can be treated as two independent modes, hence annihilation and creation operators can be assigned separately. Unlike two-dimensional dS where simply $\varphi(t,x)=\varphi(\overline{u})+\varphi(\overline{v})$ thanks to conformal flatness~\cite{Markkanen:2017abw, Aalsma:2019rpt}, four-dimensional dS does not enjoy such a luxury and we expect that the separated modes depend not simply on one of $\overline{u}$ and $\overline{v}$ exclusively, but on both of them in a mixed manner. Nevertheless, as can be read from \eqref{eq:regI-asymptoticsol1} and \eqref{eq:regI-asymptoticsol2}, around the horizon the separated modes depend only one of $\overline{u}$ and $\overline{v}$ as two-dimensional case.

In this section, we explicitly calculate the components of the energy-momentum tensor around the horizon for a minimally coupled ($\xi=0$) massless ($m=0$) scalar field,
\begin{align}
\label{eq:varphi}
\varphi(x)
\simeq
\sum_{\ell, m} \int \frac{d\omega H}{\sqrt{4\pi \omega}} \bigg\{
&
\Big[ e^{i\delta_{\omega \ell}} (H\overline{u})^{-i\omega/H} b_{\omega\ell m}^{\rm out} 
+ e^{-i\delta_{\omega \ell}} (-H\overline{v})^{i\omega/H} b_{\omega\ell m}^{\rm in} \Big] Y_\ell^m(\theta,\phi)
\nonumber\\
&
+ \Big[ e^{-i\delta_{\omega \ell}} (H\overline{u})^{i\omega/H} b_{\omega\ell m}^{{\rm out}, \dagger} 
+ e^{i\delta_{\omega \ell}} (-H\overline{v})^{-i\omega/H} b_{\omega\ell m}^{{\rm in}, \dagger} \Big]
Y_\ell^{m *}(\theta,\phi)
\bigg\}
\, ,
\end{align}
where the standard commutation relations are satisfied:
\begin{equation}
\Big[ b_{\omega\ell m}^{\rm out}, b_{\omega'\ell' m'}^{{\rm out}, \dagger} \Big]
= \Big[ b_{\omega\ell m}^{\rm in}, b_{\omega'\ell' m'}^{{\rm in}, \dagger} \Big]
= \delta(\omega-\omega')\delta_{\ell \ell'}\delta_{m m'}
\, , 
\text{ otherwise zero} 
\, .
\end{equation}
Given the components of the energy-momentum tensor in this case as
\begin{equation}
T_{\mu\nu} = \nabla_\mu\varphi \nabla_\nu\varphi - \frac12 g_{\mu\nu} (\nabla \varphi)^2 \, ,
\end{equation}
we obtain the vacuum expectation values $\langle T_{\mu\nu} \rangle$ around the horizon as follows.

While a good textbook example to study the quantum effects in curved space-time, a massless scalar field is more than a test field. For example, consider the tensor perturbations $h_{ij}$ that describe the transverse and traceless degrees of freedom in the spatial component of the Robertson-Walker metric\footnote{
One might be tempted to argue the scalar degree of freedom, the curvature perturbation, in a similar manner as it can be regarded as a scalar field with a time-dependent light mass. But it seems to bring caution in naively transforming between the static and flat coordinates \eqref{eq:flatcoord} that the curvature perturbation corresponds to the Goldstone boson of the spontaneously broken time translational symmetry in terms of the flat coordinates~\cite{Cheung:2007st}.
} [see \eqref{eq:flatcoord}]. Decomposing $h_{ij}$ as $h_{ij} = \sum_\lambda h_\lambda e_{ij}(\lambda)$ in terms of the two polarization tensors $e_{ij}(\lambda)$ with $\lambda$ being the polarization index, the quadratic action for the tensor perturbations is written covariantly as~\cite{Prokopec:2010be, Gong:2016qpq}
\begin{equation}
S_2^{(t)} = \sum_\lambda \frac{\mpl^2}{2} \int d^4x \sqrt{-g} 
\frac{-1}{2} g^{\mu\nu} \partial_\mu h_\lambda \partial_\nu h_\lambda
\, .
\end{equation} 
Other than the overall constant factor $\mpl^2/2$, which does not affect the equation of motion, each polarization mode has exactly the same action as that for a minimally coupled massless scalar field, satisfying the same equation of motion: $\square h_\lambda = 0$. Thus all the discussions here are directly applicable to the tensor perturbations, viz. primordial gravitational waves, presumably responsible for the large-scale $B$-mode polarization of the cosmic microwave background~\cite{Seljak:1996gy}.

Before we begin, it is illustrative to check the validity of our approach against the well-known textbook result $\langle\varphi^2\rangle$~\cite{Linde:1982uu,Starobinsky:1982ee,Vilenkin:1982wt} (see also~\cite{Starobinsky:1979ty}). From \eqref{eq:varphi}, we can obtain 
\begin{align}
\langle\varphi^2\rangle
& \approx 
2 \int \frac{d\omega}{4\pi\omega} H^2 \Bigg[ \sum_{\ell,m} |Y_\ell^m(\theta,\phi)|^2 \Bigg]
\nonumber\\
& \quad
+
2 \sum_{\ell,m} \int \frac{d\omega d\omega'}{4\pi\sqrt{\omega\omega'}} H^2 |Y_\ell^m(\theta,\phi)|^2
\big\langle b_{\omega'\ell m}^{\text{out},\dag}b_{\omega\ell m}^\text{out} \big\rangle
+
2 \sum_{\ell,m} \int \frac{d\omega d\omega'}{4\pi\sqrt{\omega\omega'}} H^2 |Y_\ell^m(\theta,\phi)|^2
\big\langle b_{\omega'\ell m}^{\text{in},\dag}b_{\omega\ell m}^\text{in} \big\rangle
\, .
\end{align}
If we concentrate on the first term, we can note that each in- and out-mode has a contribution proportional to $H^2$. Especially, the s-wave contribution ($\ell=0$) reads, using $Y_0^0 = 1/\sqrt{4\pi}$,
\begin{equation}
2 \int \frac{d\omega}{4\pi\omega}H^2 \times \frac{1}{4\pi}
=
2 \bigg( \frac{H}{4\pi} \bigg)^2 \log\omega
\, .
\end{equation}
That is, in each logarithmic interval of $\omega$, there is a precisely equal power $H/(4\pi)$ for each mode. This is the very well-known result for the massless scalar field in dS. Meanwhile, the other terms could be divergent depending on the vacuum choice which however can be renormalized as we will argue soon.

\subsubsection{$T_{\overline{u}\overline{u}}$ and $T_{\overline{v}\overline{v}}$}

It is straightforward to obtain
\begin{equation}
\langle T_{\overline{u}\overline{u}} \rangle 
= 
\langle \partial_{\overline u}\varphi \partial_{\overline u}\varphi \rangle 
\simeq \sum_{\ell, m} \int d\omega  \frac{\omega}{4\pi} \frac{1}{{\overline u}^2}
\bigg[ 1 + 2 \int d\omega' \big\langle b_{\omega\ell m}^{{\rm out}, \dagger} 
b_{\omega'\ell m}^{{\rm out}} \big\rangle \bigg] |Y_\ell^m(\theta, \phi)|^2
\, ,
\end{equation}
where the irrelevant oscillating terms which vanish on average over some interval of $\overline{u}$ are ignored. More concretely, the oscillating terms appear as $e^{\pm 2 i \omega u}$, which correspond to the extremely rapid oscillation for $\overline{u}=0$ (or $u=-\infty$), thus they are averaged out on ${\cal H}^-$. On ${\cal H}^+$, $t_s$ goes to infinity, so when we consider $r_s=H^{-1}(1-\epsilon)$, i.e., the region slight inside ${\cal H}^+$, the oscillation is sufficiently rapid to be averaged out. Otherwise, we may restrict out attention on the region of sufficiently large $u$ on ${\cal H}^+$, as considered in \cite{Unruh:1976db}. Since the first term corresponds to the divergent vacuum energy, we need to renormalize it. As in the Schwarzschild space-time, we expect that $\langle H | T_{\mu\nu} |H\rangle$ to be regular for a freely falling observer on both the past and  future horizons. This suggests the renormalizaton scheme $\langle  T_{\mu\nu}\rangle^{\rm ren}= \langle  T_{\mu\nu} \rangle -\langle H | T_{\mu\nu} |H\rangle$~\cite{Candelas:1980zt}. We also note that it is not explicitly summed over $\ell$ and $m$ which results in the Dirac delta function. It is because first we may be interested in the contribution from each $(\ell,m)$ mode, and second the above expression is just the limit around the horizon whereas another $\ell$ dependence appears as we move off from the horizon. In this case, only the summation over $m$ is meaningful which gives $\sum_m |Y_\ell^m|^2=(2\ell+1)/(4\pi)$.

For the Hartle-Hawking vacuum, \eqref{eq:Bogolfin} gives
\begin{align}
\langle H | T_{\overline{u}\overline{u}} |H\rangle 
& \simeq 
\sum_{\ell, m} \int \frac{d\omega}{2\pi \overline{u}^2} \frac{\omega}{e^{2\pi\omega/H} - 1} 
|Y_\ell^m(\theta, \phi)|^2 + ({\rm divergence})
\nonumber\\
& =
\sum_{\ell, m} \frac{H^2}{48\pi \overline{u}^2} |Y_\ell^m(\theta, \phi)|^2 + ({\rm divergence})
\, ,
\end{align}
and $\langle U' | T_{\overline{u}\overline{u}} |U'\rangle$ also has the same value. After renormalization, we obtain simply
\begin{equation}
\label{eq:uuHU'}
\langle H | T_{\overline{u}\overline{u}} |H\rangle^{\rm ren}
= 
\langle U' | T_{\overline{u}\overline{u}} |U'\rangle^{\rm ren}
\simeq 
0 \, .
\end{equation}
For the Boulware and the Unruh vacua, only the divergent term remains since $\langle B| b_{\omega\ell m}^{{\rm out}\dagger} b_{\omega'\ell m}^{{\rm out}} |B\rangle=\langle U| b_{\omega\ell m}^{{\rm out}\dagger} b_{\omega'\ell m}^{{\rm out}} |U\rangle=0$. After renormalization which subtracts $\langle H | T_{\overline{u}\overline{u}} |H\rangle^{\rm ren}$, we have 
\begin{equation}
\label{eq:uuBU}
\langle B | T_{\overline{u}\overline{u}} |B\rangle^{\rm ren}
= 
\langle U | T_{\overline{u}\overline{u}} |U\rangle^{\rm ren} 
\simeq 
- \sum_{\ell, m}\frac{H^2}{48\pi \overline{u}^2} |Y_\ell^m(\theta, \phi)|^2
\, .
\end{equation}
The calculation of $\langle T_{\overline{v}\overline{v}} \rangle$ can be done in the same way, so we just list the results:
\begin{equation}
\label{eq:vv}
\begin{split}
\langle H | T_{\overline{v}\overline{v}} |H\rangle^{\rm ren}
=
\langle U | T_{\overline{v}\overline{v}} |U\rangle^{\rm ren}
& \simeq 
0 \, ,
\\
\langle B | T_{\overline{v}\overline{v}} |B\rangle^{\rm ren}
= 
\langle U' | T_{\overline{v}\overline{v}} |U'\rangle^{\rm ren} 
& \simeq 
- \sum_{\ell, m}\frac{H^2}{48\pi \overline{v}^2} |Y_\ell^m(\theta, \phi)|^2
\, .
\end{split}
\end{equation}

\subsubsection{$T_{\overline{u}\overline{v}}$}

In two-dimensional dS, the trace anomaly $\langle T^\mu{}_\mu \rangle \equiv \langle T \rangle = H^2/(12\pi)$ fixes $T_{\overline{u}\overline{v}}$ to be $-H^2/[12\pi^2 (1-H^2\overline{u}\overline{v})^2]$~\cite{Davies:1976ei}. This does not happen, however, in four-dimensional dS. Even in the conformal gravity, while $\langle T\rangle = 0$, the non-zero angular components $\langle T_{\theta\theta}\rangle$ and $\langle T_{\phi\phi}\rangle$ do not fix $T_{\overline{u}\overline{v}}$ to some specific value. In the current case, we have
\begin{align}
\langle T_{\overline{u}\overline{v}}\rangle 
& = 
\frac{H^2}{(1+H^2\overline{u}\overline{v})^2}
\bigg[ (\partial_\theta\varphi)^2 +\frac{1}{\sin^2\theta} (\partial_\phi\varphi)^2 \bigg]
\nonumber\\
& \simeq 
\sum_{\ell, m} \frac{H^4}{(1+H^2\overline{u}\overline{v})^2} 
\int \frac{d\omega}{4\pi \omega} \bigg[ 2 + 4 \int d\omega'  \Big( 
\big\langle b_{\omega\ell m}^{{\rm out}, \dagger} b_{\omega'\ell m}^{{\rm out}} \big\rangle 
+ \big\langle b_{\omega\ell m}^{{\rm in}, \dagger} b_{\omega'\ell m}^{{\rm in}} \big\rangle 
\Big) \bigg] 
|\pmb{\Psi}_{\ell, m}(\theta, \phi)|^2
\, ,
\end{align} 
where we used the vector spherical harmonics, $\pmb{\Psi}_{\ell, m}(\theta, \phi) = \big( \pmb{\hat\theta}\partial_\theta + {\pmb{\hat\phi}}\sin^{-1}\theta \partial_\phi \big) Y_\ell^m$~\cite{morse-feshback}. Using $4\pi \sum_m Y_\ell^m(\theta, \phi) Y_\ell^{m *}(\theta', \phi') = (2\ell+1) P_\ell (x)$ with $x = \cos\theta \cos\theta' + \sin\theta \sin\theta' \cos(\phi-\phi')$, we find that the angular part is simplified as
\begin{equation}
\sum_{m=-\ell}^\ell |\pmb{\Psi}_{\ell, m}(\theta, \phi)|^2
=
\frac{2\ell+1}{4\pi} \lim_{(\theta', \phi')\to (\theta, \phi)}
\bigg[ \partial_\theta \partial_{\theta'} + \frac{1}{\sin\theta\sin\theta'} \partial_\phi\partial_{\phi'} \bigg] P_\ell(x) 
=
\frac{2\ell+1} {4\pi} \ell (\ell+1) 
\, .
\end{equation}
Indeed, the $\theta$- and $\phi$-derivative terms equally contribute to the summation, which reflects isotropy. 
On the other hand, the integration over $\omega$ is infrared divergent. Thus, for $|H\rangle$, we obtain
\begin{equation}
\langle H| T_{\overline{u}\overline{v}} |H\rangle 
\simeq 
\sum_\ell \frac{H^4}{8\pi^2} \ell (\ell+1) (2\ell+1)
\bigg[ \log \bigg( \frac{\varepsilon}{\Lambda} \bigg) + \frac{1}{\pi} \frac{H}{\varepsilon} \bigg]
\, ,
\end{equation}
where $\varepsilon$ and $\Lambda$ are respectively the infrared and ultraviolet cutoffs. Under the same renormalization scheme as $\langle T_{\overline{u}\overline{u}}\rangle$, i.e., subtracting the value in the Hartle-Hawking vacuum, we can set $\langle H| T_{\overline{u}\overline{v}}|H\rangle =0$. This, however, does not get rid of the infrared divergence of $\langle T_{\overline{u}\overline{v}}\rangle$ in other vacua completely:
\begin{align}
\langle B| T_{\overline{u}\overline{v}} |B\rangle^{\rm ren} 
& \simeq 
- \sum_\ell \frac{H^4}{8\pi^2} \ell (\ell+1) (2\ell+1) \frac{1}{\pi} \frac{H}{\varepsilon}
\, ,
\\
\langle U| T_{\overline{u}\overline{v}} |U\rangle^{\rm ren} 
= 
\langle U'| T_{\overline{u}\overline{v}} |U'\rangle^{\rm ren} 
& \simeq 
- \sum_\ell \frac{H^4}{16\pi^2} \ell (\ell+1) (2\ell+1) \frac{1}{\pi} \frac{H}{\varepsilon}
\, .
\end{align}

\subsubsection{Angular components}

The angular components of the energy-momentum tensor are given by
\begin{align}
\langle T_{\theta\theta}\rangle
& =
\frac12 \bigg[ \langle \partial_\theta\varphi \partial_\theta\varphi \rangle 
- \frac{1}{\sin^2\theta} \langle \partial_\phi\varphi \partial_\phi\varphi \rangle \bigg]
+ \frac{(1+H^2\overline{u}\overline{v})^2}{2H^2} 
\langle \partial_{\overline u}\varphi \partial_{\overline v}\varphi \rangle
\, ,
\\
\frac{\langle T_{\phi\phi}\rangle}{\sin^2\theta}
& = 
\frac12 \bigg[ \frac{1}{\sin^2\theta} \langle \partial_\phi\varphi \partial_\phi\varphi \rangle
- \langle \partial_\theta\varphi \partial_\theta\varphi \rangle \bigg]
+ \frac{(1+H^2\overline{u}\overline{v})^2}{2H^2}
\langle \partial_{\overline u}\varphi \partial_{\overline v}\varphi \rangle
\, ,
\\
\langle T_{\theta\phi} \rangle
& = 
\langle \partial_\theta\varphi \partial_\phi\varphi\rangle
\, .
\end{align}
Indeed, these three terms vanish around the horizon, because
\begin{align}
\sum_{m=-\ell}^\ell |\partial_\theta Y_\ell^m(\theta, \phi)|^2
& =
\frac{2\ell+1}{8\pi} \ell (\ell+1)
\, ,
\\
\sum_{m=-\ell}^\ell \frac{1}{\sin^2\theta} |\partial_\phi Y_\ell^m(\theta, \phi)|^2
& =
\frac{2\ell+1}{8\pi} \ell (\ell+1)
\, ,
\\
\sum_{m=-\ell}^\ell \partial_\theta Y_\ell^m(\theta, \phi) \partial_\phi Y_\ell^{m*}(\theta, \phi)
& =
0 \, ,
\end{align}
which reflects isotropy, and also because $\langle \partial_{\overline u}\varphi\partial_{\overline v}\varphi \rangle \simeq 0$ which follows from the separation of the modes around the horizon.

\subsection{Energy-momentum tensor in other coordinates}
\label{sec:othercoor}

While the Kruskal-Szekeres null coordinates are useful to discuss the separation of $\overline{u}$- and $\overline{v}$-dependent modes around the horizon, it is instructive to investigate the components of the energy-momentum tensor in other coordinate systems.

\subsubsection{Tortoise coordinates}

The tortoise coordinates are related to the Kruskal-Szekeres null coordinates through \eqref{eq:statKS}.
The energy-momentum tensor components in terms of the tortoise coordinates are given by
\begin{equation}
\label{eq:Tstatic}
\begin{split}
\langle T_{t_st_s} \rangle 
& =
e^{2H(t_s-r_*)} \langle T_{\overline{u}\overline{u}} \rangle
+ e^{-2H(t_s+r_*)} \langle T_{\overline{v}\overline{v}} \rangle
+ 2 e^{-2Hr_*} \langle T_{\overline{u}\overline{v}} \rangle
\, ,
\\
\langle T_{r_*r_*} \rangle 
& =
e^{2H(t_s-r_*)} \langle T_{\overline{u}\overline{u}} \rangle
+ e^{-2H(t_s+r_*)} \langle T_{\overline{v}\overline{v}} \rangle
- 2e^{-2Hr_*} \langle T_{\overline{u}\overline{v}} \rangle
\\
\langle T_{t_sr_*} \rangle 
& =
- e^{2H(t_s-r_*)} \langle T_{\overline{u}\overline{u}} \rangle
+ e^{-2H(t_s+r_*)} \langle T_{\overline{v}\overline{v}} \rangle
\, .
\end{split}
\end{equation}
Now we consider how they in different vacua behave around the horizon, referring to \eqref{eq:uuHU'}, \eqref{eq:uuBU}, and \eqref{eq:vv}. First, all the components in the Hartle-Hawking vacuum vanish after renormalization. In the Boulware vacuum, we have non-zero renormalized energy-momentum tensor components in which the contributions from $\langle T_{\overline{u}\overline{u}}\rangle$ and $\langle T_{\overline{v}\overline{v}}\rangle$ are the same:
\begin{align}
e^{2H(t_s-r_*)} \langle T_{\overline{u}\overline{u}} \rangle
& \simeq 
- e^{2H(t_s-r_*)} \sum_{\ell, m} \frac{H^2}{48\pi \overline{u}^2} |Y_\ell^m(\theta, \phi)|^2
= 
- \sum_{\ell, m} \frac{H^2}{48\pi (1/H)^2} |Y_\ell^m(\theta, \phi)|^2
\, ,
\\
e^{-2H(t_s+r_*)} \langle T_{\overline{v}\overline{v}}\rangle 
& \simeq 
- e^{-2H(t_s+r_*)} \sum_{\ell, m} \frac{H^2}{48\pi \overline{v}^2} |Y_\ell^m(\theta, \phi)|^2
= 
- \sum_{\ell, m} \frac{H^2}{48\pi (1/H)^2} |Y_\ell^m(\theta, \phi)|^2
\, .
\end{align}
This indicates 
\begin{align}
\langle H | T_{t_sr_*} |H\rangle = \langle B | T_{t_sr_*} |B\rangle
& = 0
\, ,
\\
\langle U | T_{t_sr_*} |U\rangle 
& \simeq 
\sum_{\ell, m} \frac{H^2}{48\pi (1/H)^2} |Y_\ell^m(\theta, \phi)|^2
\, ,
\\
\langle U' | T_{t_sr_*} |U'\rangle 
& \simeq  
- \sum_{\ell, m} \frac{H^2}{48\pi (1/H)^2} |Y_\ell^m(\theta, \phi)|^2
\, .
\end{align}
The physical meaning of $\langle T_{t_sr_*} \rangle$ becomes clear when we define the ``luminosity'' $L$ at $r_s$ by $\langle  T_{t_sr_*} \rangle=-L/(4\pi r_s^2)$~\cite{Schutz:1985jx}, which is interpreted as the flux of thermal energy that escapes from or entering into the surface of constant $r_s$. Therefore, we find that for the Hartle-Hawking and the Boulware vacua, thermal radiation at the horizon $r_s=H^{-1}$ does not give rise to any net energy flow going into or out of the horizon. This in fact can be intuitively understood from the effective potential \eqref{eq:effpot} (see also Figure~\ref{fig:dSwave}). Since the wave from ${\cal H}^-$ is not transmitted to $r_s=0$ but reflected to ${\cal H}^+$ only, the reflection rate is exactly $1$, which implies the balance between the thermal fluxes incoming from ${\cal H}^-$ and outgoing to ${\cal H}^+$. The same absolute value of coefficients for the incoming and outgoing waves in \eqref{eq:horwave} shows this fact explicitly. Only the Unruh vacua -- $|U\rangle$ and $|U'\rangle$ -- allow a non-zero flux resulting from thermal radiation, in which case the luminosity at $r_s=H^{-1}$ is given by, for each $(\ell, m)$ mode, 
\begin{equation}
\label{eq:luminosity}
L = \mp \frac{H^2}{12} |Y_\ell^m|^2 
\text{ for } 
|U\rangle \text{ and } |U'\rangle 
\, .
\end{equation}

We also compare the components of the energy-momentum tensor we obtained with those in~\cite{Aalsma:2019rpt}. From general covariance, we expect that the renormalized energy-momentum tensor is conserved, i.e., $\nabla^\mu\langle T_{\mu\nu}\rangle^{\rm ren} = 0$ (for other conditions that $\langle T_{\mu\nu}\rangle^{\rm ren}$ should satisfy, see~\cite{Wald:1977up}). If $\langle T_{\mu\nu}\rangle^{\rm ren}$ depend only on $r_s$, this condition in the static coordinates gives 
\begin{align}
\partial_{r_s} \langle T^{t_s r_s} \rangle
& =
\frac{2(2H^2r_s^2-1)}{r_s(1-H^2r_s^2)} \langle T^{t_s r_s} \rangle
\, ,
\\
\partial_{r_s} \langle T^{r_s r_s} \rangle
& =
- \frac{2}{r_s} \langle T^{r_s r_s} \rangle + 2 r_s \langle T^{\theta\theta} \rangle - H^2 r_s \langle T \rangle
\, ,
\\
\langle T^{\phi\phi} \rangle
& =
\sin^{-2}\theta \langle T^{\theta\theta} \rangle
\, .
\end{align}
These are solved to give~\cite{Aalsma:2019rpt}
\begin{align}
\langle T_{t_s t_s} \rangle
& =
\frac{1}{r_s^2} \bigg\{ \Delta - H^2 \int_{1/H}^{r_s} dr {r}^3 \langle T(r) \rangle + \Theta(r_s)
+ \big( 1 - H^2r_s^2 \big) \Big[ 2 \langle T_{\theta\theta}(r_s) \rangle - r^2 \langle T(r_s) \rangle \Big] \bigg\}
\, ,
\\
\langle T_{r_* r_*} \rangle
& = 
\frac{1}{r_s^2} \bigg[ \Delta - H^2 \int_{1/H}^{r_s} dr {r}^3 \langle T(r) \rangle + \Theta(r_s) \bigg]
\, ,
\\
\langle T_{t_s r_*} \rangle
& =
- \frac{\Phi}{r_s^2} 
\, ,
\end{align}
where $\Delta$ and $\Phi$ are integration constants, and
\begin{equation}
\Theta(r_s) = 2 \int_{1/H}^{r_s} dr \frac{\langle T_{\theta\theta}(r) \rangle}{r} \, .
\end{equation}
They can be applied to our case for the s-wave mode, which does not depend on the angular variables. In this case the trace
\begin{equation}
\langle T \rangle
=
\big( 1 - H^2\overline{u}\overline{v} \big)^2 
\langle \partial_{\overline u}\varphi \partial_{\overline v}\varphi \rangle
- H^2 \bigg( \frac{1-H^2\overline{u}\overline{v}}{1+H^2\overline{u}\overline{v}} \bigg)^2
\Big\langle (\partial_\theta \varphi)^2 + \sin^{-2}\theta (\partial_\phi\varphi)^2 \Big\rangle
\end{equation}
vanishes around the horizon due to 1) the separation of $\overline{u}$ and $\overline{v}$ modes around the horizon, and 2) the exclusive dependence on $r_s$. Then only the terms involving the integration constants survive in $\langle T_{\mu\nu} \rangle$ satisfying $\langle T_{t_s t_s} \rangle = \langle T_{r_* r_*} \rangle$. This is consistent with \eqref{eq:Tstatic}, as $T_{\overline{u}\overline{v}}=0$ for $\ell=0$ and $r_s=H^{-1}$. We also note that the luminosity corresponds to the integration constant, $L=4\pi\Phi$.

\subsubsection{Flat coordinates}
\label{sec:flatcoord}

For cosmological applications, the flat coordinates are useful in which the metric is given by
\begin{equation}
\label{eq:flatcoord}
ds^2 = \frac{1}{H^2\tau^2} \Big[ -d\tau^2 + dr^2 + r^2 \big( d\theta^2 + \sin^2\theta d\phi^2 \big) \Big] \, .
\end{equation}
Here we consider the conformal time $\tau$ instead of the usual flat time coordinate $t$, related by $\tau=-H^{-1}e^{-Ht}$, since $\tau$ is directly connected to the Kruskal-Szekeres coordinates through
\begin{equation}
\label{eq:KSflat}
H\overline{u} = - \frac{1}{H(\tau-r)}
\quad \text{and} \quad 
H\overline{v} = H(\tau+r)
\, .
\end{equation}
This implies that the positive modes with respect to $\overline{u}$ and $\overline{v}$ are just those with respect to $\tau$. This is the reason why the Hartle-Hawking vacuum in the static coordinates corresponds to the Bunch-Davies vacuum. The components of the energy momentum tensor in the flat coordinates are written as
\begin{align}
\langle T_{\tau\tau} \rangle 
& =
\frac{1}{H^4(\tau-r)^4} \langle T_{\overline{u}\overline{u}} \rangle
+ \langle T_{\overline{v}\overline{v}} \rangle
+ \frac{2}{H^2(\tau-r)^2} \langle T_{\overline{u}\overline{v}} \rangle
\, ,
\\
\langle T_{rr} \rangle 
& =
\frac{1}{H^4(\tau-r)^4} \langle T_{\overline{u}\overline{u}} \rangle 
+ \langle T_{\overline{v}\overline{v}} \rangle 
- \frac{2}{H^2(\tau-r)^2} \langle T_{\overline{u}\overline{v}}\rangle
\, ,
\\
\langle T_{\tau r} \rangle 
& =
- \frac{1}{H^4(\tau-r)^4} \langle T_{\overline{u}\overline{u}} \rangle
+ \langle T_{\overline{v}\overline{v}} \rangle
\, .
\end{align}
These are in agreement with~\cite{Markkanen:2017abw}.

We note that whereas the static coordinates only cover Region I, the flat coordinates cover Region I and II, which includes ${\cal H}^+$. Since $\langle T_{\overline{u}\overline{u}}\rangle$ for $|B\rangle$ and $|U\rangle$  diverges as $1/\overline{u}^2$ around ${\cal H}^{-}$ ($\overline{u}=0$) and $\langle T_{\overline{v}\overline{v}}\rangle$  for $|B\rangle$ and $|U'\rangle$ diverges as $1/\overline{v}^{2}$ around ${\cal H}^{+}$ ($\overline{v}=0$), we find that $|B\rangle$ and $|U'\rangle$ are not well-defined in the flat coordinates~\cite{Aalsma:2019rpt}. Nevertheless, we restrict our attention to Region I and investigate thermal flux and the backreaction in $|B\rangle$ and $|U'\rangle$ as well, since any observer moving along the time-like trajectory cannot see what happens in Region II in the past before dS is significantly deformed.

In the Boulware vacuum,
\begin{align}
\frac{1}{H^4(\tau-r)^4} \langle T_{\overline{u}\overline{u}} \rangle
& =
- \frac{1}{H^4(\tau-r)^4} \sum_{\ell, m} \frac{H^2}{48\pi \overline{u}^2} |Y_\ell^m(\theta, \phi)|^2 
=
- \frac{1}{(\tau-r)^2} \sum_{\ell, m} \frac{H^2}{48\pi} |Y_\ell^m(\theta, \phi)|^2
\, ,
\\
\langle T_{\overline{v}\overline{v}} \rangle
& =
\sum_{\ell, m} \frac{H^2}{48\pi \overline{v}^2} |Y_\ell^m(\theta, \phi)|^2 
=
- \frac{1}{(\tau+r)^2} \sum_{\ell, m} \frac{H^2}{48\pi} |Y_\ell^m(\theta, \phi)|^2
\, ,
\end{align}
which indicate that the cancellation between the two contributions does not occur in $\langle T_{\tau r} \rangle$. Such a non-zero $\langle T_{\tau r} \rangle$ may give rise to, for example, $g_{\tau r}$ through backreaction. Indeed, the luminosity in which the cancellation in thermal flux is evident is defined in the static coordinates, where the horizon is given by a constant radial coordinate, $r_s=H^{-1}$. But the horizon is not static in the flat coordinates, as can be seen from the relations
\begin{equation}
\label{eq:statflat}
r_s = r e^{Ht}
\quad \text{and} \quad 
e^{-Ht_s} = e^{-Ht} \sqrt{1-H^2 r^2 e^{2Ht}}
\, ,
\end{equation}
which makes the physical interpretation of a non-zero $\langle T_{\tau r} \rangle$ unclear when we focus on what happens around the horizon.

Another ambiguity also arises in the Unruh vacuum. A negative (positive) luminosity for $|U\rangle$ ($|U'\rangle$) has an obvious interpretation that the energy flux resulting from thermal radiation is absorbed into (emitted from) the horizon. To see this, we note that since the energy density in the flat coordinates is given by $\rho=3\mpl^2 H^2$ while the volume that corresponds to the horizon size $r_s = H^{-1}$, the ``comoving'' volume with the radius $r=H^{-1}e^{-Ht}$, is $V=(4\pi/3)(H^{-1}e^{-Ht})^3$, the total physical energy inside this volume becomes $E=a^3\rho V=4\pi \mpl^2 /H$. Then we expect that in $|U\rangle$, where the outgoing mode contributes to the flux as reflected in the negative luminosity, the energy inside the horizon decreases in time. This energy is transferred to the horizon hence $H$ increases in time, or equivalently the horizon volume decreases. In the same way, we also expect that in $|U'\rangle$, where the incoming mode contributes to the flux, $H$ decreases in time. However, as 
\begin{equation}
\rho+p = H^2\tau^2(\langle T_{\tau\tau} \rangle+\langle T_{rr} \rangle) = -2\mpl^2\dot{H} < 0 
\end{equation}
for both $|U\rangle$ and $|U'\rangle$, it seems that unlike the expectation, backreaction {\it always} makes $H$ increase in time.

\section{Change in the horizon area by backreaction}
\label{sec:area}

The discussion in the previous section suggests that instead of naively considering the Friedmann equation, we need to see the change in the horizon area from the geometric point of view. The change in the horizon area along the null geodesic, whose tangent vector is normal as well as tangent to the horizon simultaneously, is parametrized by the expansion $\Theta$. While its definition and properties can be found in standard textbooks, e.g.,~\cite{Wald:1984rg}, we briefly summarize them in Appendix~\ref{app:Killhor} to establish notations and conventions as well to render our discussion self-contained. Only in the following paragraph and in Appendix~\ref{app:Killhor}, we use the abstract index notation.

To begin with, we consider a vector field $U^a$ normal to the event horizon. Since the horizon is a null hypersurface, $U^a$ is null so it is also tangent to the horizon such that we can find the geodesic congruence on the horizon which is tangent to $U^a$. Denoting the affine parameter for this geodesic congruence as $\lambda$, the change in the horizon area $\calA$ along the geodesic is given by~\cite{Jacobson:1995ab} (see also Appendix~\ref{app:area} for justification)
\begin{equation}
\frac{d}{d \lambda}{\cal A} = \int \Theta d{\cal A} \, .
\end{equation}
Here, $\Theta$ is called the expansion, which measures the rate at which the deviation between nearby geodesics generated by $U^a$ expands or shrinks along $\lambda$. The evolution of $\Theta$ is determined by the energy-momentum tensor, as well as the shear $\widehat{\sigma}_{ab}$ and the rotation $\widehat{\omega}_{ab}$ of the deviation between nearby geodesics, as given by the Raychaudhuri equation (see Appendix~\ref{app:Rayeq}):
\begin{equation}
\label{eq:Ray}
\frac{d\Theta}{d\lambda}
= 
- \frac12 \Theta^2 - \widehat{\sigma}^{ab}\widehat{\sigma}_{ab} 
+ \widehat{\omega}^{ab}\widehat{\omega}_{ab} - \frac{1}{\mpl^2} T_{ab}U^aU^b
\, .
\end{equation}
On the other hand, we learn from discussion below \eqref{eq:dSkruskal} that $U^a$ of the dS event horizon is in fact the time-like Killing vector $\partial_{t_s}$ so the dS horizon becomes the Killing horizon. Then it is known that  $\Theta=\widehat{\sigma}_{ab}=\widehat{\omega}_{ab}=0$ on the dS horizon (see Appendix~\ref{app:vanKill}). Indeed, on ${\cal H}^+$ (${\cal H}^-$), since $U^a$ is parallel to $k^a=H\overline{u}(\partial_{\overline u})^a=(H\overline{u})^{-1}(\partial_\tau-\partial_r)^a$ or equivalently $k^a=-H\overline{v}(\partial_{\overline v})^a = -H\overline{v}(\partial_\tau+\partial_r)^a$, the energy-momentum tensor contribution $\mpl^{-2} T_{ab}U^aU^b$ in \eqref{eq:Ray} is proportional to $T_{\tau\tau} + T_{rr}$, or $\rho+p$. Thus in the absence of backreaction, $\Theta$, $\widehat{\sigma}_{ab}$, and $\widehat{\omega}_{ab}$ vanish identically on the horizon. Then the horizon area ${\cal A}=4\pi/H^2$ is a constant.

Note that usually the last term of \eqref{eq:Ray} is written in terms of the Ricci tensor as in \eqref{eq:raychaudhuri}, so in fact the Raychaudhuri equation is purely geometric. However, in this article we are interested in how the matter backreaction (currently thermal radiation of a minimally coupled massless scalar field) modifies the geometry via the semi-classical Einstein equation \eqref{eq:einstein}. Thus the Ricci tensor is replaced by the energy-momentum tensor so that we can observe directly the effects of thermal radiation on the geometry $\Theta$.

To see how the backreaction from thermal radiation modifies the situation, we first consider the future horizon ${\cal H}^+$, on which $d/d\lambda=d/d{\overline u}$, or $U^a=(\partial_{\overline u})^a$ is taken. For $|B\rangle$ or $|U \rangle$, the Raychaudhuri equation becomes
\begin{equation}
\label{eq:Rayu}
\frac{d\Theta}{d{\overline u}} = - \frac{T_{{\overline u}{\overline u}}}{\mpl^2}
= \sum_{\ell, m} \frac{H^2}{48\pi \mpl^2 {\overline u}^2} |Y_\ell^m(\theta, \phi)|^2
\, ,
\end{equation}
since $\Theta=\widehat{\sigma}_{ab}=\widehat{\omega}_{ab}=0$. This is solved to give the backreacted expansion $\Theta=-\sum_{\ell, m} [H^2/(48\pi \mpl^2\overline{u})] |Y_\ell^m|^2$, from which the rate of change in the horizon area is written as
\begin{equation}
\frac{ {\overline u}}{{\cal A}} \frac{d{\cal A}}{d{\overline u}}
= 
- \frac{2}{H^2} \frac{dH}{du}
\simeq 
- \sum_{\ell, m} \frac{H^2}{48\pi \mpl^2} |Y_\ell^m(\theta, \phi)|^2
\, .
\end{equation}
Since the right-hand side is negative while $\overline{u}$ in the left-hand side is positive on ${\cal H}^+$, the horizon area decreases along the increasing direction of $\overline{u}$: $d{\cal A}/d\overline{u}<0$. Another way to see this is to notice from \eqref{eq:KSflat} and \eqref{eq:statflat} that $(H\overline{u})^{-1}=-H(\tau-r)=(1+H r_s)e^{-Ht}$. Around the horizon, the change in $\overline{u}$ comes from the flat time coordinate $t$, i.e., $d/d\overline{u}=(H\overline{u})^{-1}d/dt$, from which we find
\begin{equation}
\epsilon_H \equiv -\frac{1}{H^2}\frac{dH}{dt}
= - \sum_{\ell, m} \frac{H^2}{96\pi \mpl^2}|Y_\ell^m(\theta, \phi)|^2 
\, .
\end{equation}
Thus, in the cosmological context in which the flat time coordinate $t$ measures time, $H$ {\it increases in time} which results in the decrease in the horizon area, as also obtained in~\cite{Aalsma:2019rpt}.

A similar result can be drawn for the past horizon ${\cal H}^-$. Since $d/d\lambda=d/d\overline{v}$ in this case, we just replace $\overline{u}$ in \eqref{eq:Rayu} by $\overline{v}$ for $|B\rangle$ or $|U'\rangle$. From this, the backreacted expansion is given by $\Theta=-\sum_{\ell, m}[H^2/(48\pi \mpl^2\overline{v})]|Y_\ell^m|^2$. We note here that even though it is a mere replacement of $\overline{u}$ by $\overline{v}$, since $\overline{v}<0$ on ${\cal H}^-$, we have positive $\Theta$, the effect of which to the horizon area on ${\cal H}^-$ is opposite to that on ${\cal H}^+$. Indeed, the rate of change in the horizon area reads
\begin{equation}
\frac{ {\overline v}}{{\cal A}} \frac{d{\cal A}}{d{\overline v}}
= 
\frac{2}{H^2} \frac{dH}{dv}
\simeq  
- \sum_{\ell, m} \frac{H^2}{48\pi \mpl^2} |Y_\ell^m(\theta, \phi)|^2
\, ,
\end{equation}
which shows that the horizon area {\it increases} along the increasing direction of ${\overline v}$: $d{\cal A}/d\overline{v}>0$. To see how it appears in cosmology, we note the coordinate transformation $H\overline{v}=H(\tau+r)=-(1-H r_s)e^{-Ht}$. The problem here is that at $r_s=H^{-1}$, $\overline{v}$ becomes zero unless $t=-\infty$, which is manifest in the Penrose diagram. This seems the reason why the Raychaudhuri equation does not describe the increase in the horizon area along $\overline{v}$ on ${\cal H}^-$. The Raychaudhuri equation tells us the time evolution of $H$ along the flat time coordinate, $t$-direction\footnote{
To be precisely, \eqref{eq:ray-FRW} is obtained by combining the Raychaudhuri equation and the energy constraint. With $a(t)$ being the scale factor in the Friedmann-Robertson-Walker metric, the Raychaudhuri equation alone reads
\begin{equation*}
\frac{\ddot{a}}{a} = - \frac{\rho+3p}{6\mpl^2} \, .
\end{equation*}
}:
\begin{equation}
\label{eq:ray-FRW}
\dot{H} = - \frac{\rho+p}{2\mpl^2} \, ,
\end{equation}
where a dot denotes a derivative with respect to the flat time coordinate. Since the hypersurface $t=$ constant intersects with ${\cal H}^+$ only once and the increasing direction of $t$ coincides with that of $\overline{u}$ on ${\cal H}^+$, we infer that the change in $H$ along the $t$-direction and that along the $\overline{u}$-direction are essentially equivalent. On the other hand, $t$ is identically given by $-\infty$ on ${\cal H}^-$ such that the change in $t$ on ${\cal H}^-$ is {\it not} reflected in the Raychaudhuri equation. Even in this case, we can consider a constant $r_s= H^{-1}(1-\epsilon)$ surface with $\epsilon \ll 1$, which is slightly deviated from ${\cal H}^-$ and identify the change in $\overline{v}$ with that in $t$ on the surface. In this case, we obtain $d/d\overline{v}=-(H\overline{v})^{-1}d/dt$ (note that since $\overline{v}<0$ on ${\cal H}^-$, the increasing direction of $\overline{v}$ coincides with that of $t$), from which we find
\begin{equation}
\label{eq:epsilon}
\epsilon_H = \sum_{\ell, m}\frac{H^2}{96\pi \mpl^2} |Y_\ell^m(\theta, \phi)|^2 \, ,
\end{equation}
therefore $H$ {\it decreases} in the flat time coordinate.

For $|B\rangle$, the changes in the horizon area  arise on ${\cal H}^+$ and ${\cal H}^-$ in the same magnitude but opposite sign such that these two contributions are cancelled with each other. As a result, the horizon area does not change in time. It is consistent with the feature discussed in Section~\ref{sec:othercoor} that $|B\rangle$ does not have a net thermal flux as indicated by the vanishing luminosity, hence the region inside the horizon does not lose or gain energy. On the other hand, in $|U\rangle$, only the thermal flux on ${\cal H}^+$ survives, which makes the horizon area decrease in time. The opposite situation arises in $|U'\rangle$, such that only survives the thermal flux on ${\cal H}^-$, the sign of which is opposite to that on ${\cal H}^+$ and thus the horizon area increases in time. In this way, our estimation on the change in the horizon area using the expansion $\Theta$ is consistent with the behavior of thermal flux as quantified by luminosity.

In fact, such a consistency assumes adiabaticity, i.e., the horizon area changes sufficiently slowly compared to the backreaction time scale. In this case, the system is close to thermal equilibrium and the total entropy is almost constant in time. The contribution to the entropy from geometry is given by the area law, $S_{\rm dS}=8\pi^2 \mpl^2/H^2$~\cite{Gibbons:1977mu}, such that\footnote{
More precisely, since the horizon radius is written in terms of the static coordinate $r_s=H^{-1}$, the natural time must be the static time coordinate $t_s$. While $\epsilon_H$ is defined in terms of the flat time coordinate $t$, it is related to $t_s$ by $e^{-Ht_s}=e^{-Ht}\sqrt{1-H^2r_s^2}$. For a fixed static radius $r_s=H^{-1}(1-\epsilon)$, $t_s$ and $t$ are related as $t_s=t-\log(2\epsilon)/2$  then we find $dt_s=dt$.
}
\begin{equation}
\frac{d S_{\rm dS}}{dt_s} = 2 \epsilon_H S_{\rm dS} H \, .
\end{equation} 
On the other hand, the thermal flux carried by the outgoing modes enhances the entropy as our ignorance on the states beyond the horizon increases in time, while that carried by the incoming modes reduces the entropy for the opposite reason. Then the change in the entropy contributed by thermal flux is written in terms of the luminosity as
\begin{equation}
\frac{dS_{\rm rad}}{dt_s} = \frac{1}{T} \frac{dQ}{dt_s} = - \frac{2\pi}{H} \frac{L}{4\pi r_s^2} (4\pi r_s^2) \, ,
\end{equation}
where the temperature is given by the Gibbons-Hawking temperature $T=H/(2\pi)$~\cite{Gibbons:1977mu} and the minus sign in the right-hand side indicates that the outgoing (incoming) thermal flux increases (decreases) thermal energy in the region beyond the horizon, $r_s>H^{-1}$. From \eqref{eq:luminosity}, the adiabatic condition $d(S_{\rm dS}+S_{\rm rad})/dt_s \simeq 0$ reads
\begin{equation}
\epsilon_H \simeq \mp \sum_{\ell, m} \frac{H^2}{96\pi \mpl^2} |Y_\ell^m(\theta, \phi)|^2 \, ,
\end{equation}
which exactly reproduces our previous results.

Note that while we have shown the total contribution from all $\ell$ modes, naively summing them all would make $\epsilon_H$ diverging. As we have seen in Section~\ref{subsec:dSsep}, however, the barrier of the effective potential vanishes only for $\ell = 0$, so the s-wave mode can reach the observer at the origin. Then she finds the effects of thermal radiation modify the classical dS by the rate
\begin{equation}
\label{eq:epsilon-swave}
|\epsilon_H| = \frac{H^2}{384\pi^2\mpl^2} \, .
\end{equation}
This is very close to what was found previously~\cite{Markkanen:2017abw}. Meanwhile, since $R_{\omega,\ell=0}$ does not oscillate but approaches a constant at $r_s=0$, it is not appropriate to say the s-wave mode ``reaches'' the observer at $r_s=0$ as we cannot define the incoming or outgoing flux by oscillating waves.

\section{Discussions and Conclusions}
\label{sec:conclusion}

Our discussions so far visit the possibility that the backreaction induced by thermal radiation breaks the SO(1,4) dS isometry. At first glance, obviously thermal radiation can destabilize dS from the facts that 1) it does not satisfy $\rho+p=0$ and 2) in the black hole analogy, it actually becomes the origin of black hole evaporation. However, the situation is not so trivial, since space-time may be in thermal equilibrium with thermal radiation such that no actual dynamical change takes place. In the Schwarzschild space-time, this is realized when the quantum state is in the Hartle-Hawking vacuum. Indeed, black hole evaporation is modeled in the Schwarzschild geometry by imposing the imbalance between the incoming and outgoing thermal fluxes, as realized in the Unruh vacuum.

In dS, only the region corresponding to thermal atmosphere exists, and there is no counterpart to the exterior of a black hole, which has to do with the absence of the asymptotic flatness in dS. This restricts the form of the solution to the wave equation, such that the wave coming from ${\cal H}^-$ is just reflected to ${\cal H}^+$, instead of being transmitted to $r_s=0$. Since an observer at $r_s=0$ does not see the propagation of the wave, she cannot compare the positive modes at ${\cal H}^\pm$ with those at $r_s=0$ to discuss thermal properties of space-time. In this work, we have argued in the context of four-dimensional dS that even in this case  the incoming and outgoing modes can be separated  when we model the interaction between thermal radiation and ultraviolet physics that breaks the dS isometry as a scattering process. Then by assigning different positive frequencies to each mode, we can construct the Unruh vacuum in dS. Just like the black hole case, the Unruh vacuum deforms dS by expanding or shrinking the horizon, which can be explicitly quantified by the expansion $\Theta$. The expansion indeed provides the description for the change in the horizon area consistent with the energy flux of thermal radiation. The same results are obtained from the adiabaticity assumption, under which entropy is almost constant in time.

Our result $|\epsilon_H| \sim {\cal O}(10^{-3}) H^2/\mpl^2$ per each $(\ell, m)$ mode already appeared in previous literatures, e.g.,~\cite{Markkanen:2016jhg, Markkanen:2016vrp, Markkanen:2017abw, Aalsma:2019rpt}. Away from the numerical factor, this is remarkable since in slow-roll inflation $\epsilon_H > H^2/\mpl^2$ forbids eternal inflation\footnote{
When the universe is in non-trivial excited states, some integer multiple of $H^2/\mpl^2$ may allow eternal inflation~\cite{Seo:2020ger}.
}, which arises when the quantum fluctuations stretched beyond the horizon (hence becomes classical perturbations through decoherence: See Footnote~\ref{footnote:deco}) generates the probability that the inflaton fluctuates up the potential~\cite{Steinhardt:1983, Vilenkin:1983xq, Linde:1986fc, Linde:1986fd, Goncharov:1987ir} (see~\cite{Guth:2007ng} for a brief review and~\cite{Starobinsky:1982ee,Starobinsky:1986fx} for possible stochastic effects). The Unruh vacuum $|U'\rangle$, in which $H$ decreases in time, is relevant to this issue as thermal radiation induces slow-roll inflation through backreaction. In this case, even if we start from space-time close to perfect dS satisfying $\epsilon_H \approx 0$\footnote{
In perfect dS $\epsilon_H=0$, the classical solution to the equation of motion is just $\varphi=$ constant. In order that the inflaton moves in time classically, the geometry needs to be quasi-dS. In this case, the quantum fluctuation of the inflaton combines with that of the trace part of the metric in a gauge-invariant way~\cite{Mukhanov:1990me}. This new field, called the curvature perturbation previously mentioned, is in fact a unique gauge-invariant combination from quantum field theoretic point of view~\cite{Gong:2016qpq}. 
}
the backreaction from thermal radiation generates $\epsilon_H\sim {\cal O}(10^{-3})H^2/\mpl^2$. If the number of particle species or degrees of freedom specified by $(\ell, m)$ modes is large enough, their total contributions can make $\epsilon_H \sim H^2/\mpl^2$, then the universe leaves eternally inflating regime and enters the slow-roll phase of inflation. Presumably the numerical factor of ${\cal O}(10^{-3})$ attached to $H^2/\mpl^2$ together with the number of radiation degrees of freedom in effective theory may provide a window that eternal inflation can take place, whereas there can be model dependence in the numerical factor allowing eternal inflation~\cite{Rudelius:2019cfh}.

It is at this point interesting to notice that the change in $S_\text{rad}$ in a given $e$-fold duration $\Delta{N} = H\Delta{t}$ is constant:
\begin{equation}
\Delta{S}_\text{rad} = \pm \frac{\pi}{6} \sum_{\ell,m} |Y_\ell^m|^2 \, ,
\end{equation}
where the positive (negative) sign is for $|U\rangle$ ($|U'\rangle$): the entropy of thermal radiation increases (decreases) monotonically for $|U\rangle$ ($|U'\rangle$). Especially, only considering the s-wave contribution, $\Delta{S}_\text{rad} = \pm1/24$. The adiabatic equilibrium condition demands $\Delta{S}_\text{dS}$ compensate $\Delta{S}_\text{rad}$. Then $\Delta{S}_\text{dS} = \mp1/24$ for $\ell=0$, which is in sharp contrast to $\Delta{S}_\text{dS} \gg 1$ in slow-roll inflation~\cite{ArkaniHamed:2007ky}. Nevertheless, for $|U'\rangle$ we expect a definite adiabatic evolution of the dS geometry all the way as $\Delta{S}_\text{dS}>0$: Solving \eqref{eq:epsilon-swave} for $H$ gives~\cite{Padmanabhan:2002ji}
\begin{equation}
\label{eq:Hevolution}
H(t) = H_0 \bigg( 1 + \frac{1}{128\pi^2} \frac{H_0^3}{\mpl^2} t \bigg)^{-1/3} \, ,
\end{equation}
where $H_0$ is the value ofthe Hubble parameter when the dS stage begins at $t=0$, so both $H$ and $\epsilon_H$ becomes zero as $t\to\infty$. On the other hand, for $|U\rangle$ the sign for the second term in the parentheses of the above equation flips, such that $H(t)$ becomes singular when $t = 128\pi^2\mpl^2/H_0^3$. This means our semi-classical, perturbative approach to consider a (minimally coupled, massless) scalar field for the instability of dS becomes invalid after this time scale.

Indeed, given $\epsilon_H=c H^2/\mpl^2$ with some constant $c$, a significant change in the Hubble parameter occurs as $\Delta H/H \sim {\cal O}(1)$  after the ``quantum break-time'' $\Delta t \sim c^{-1}\mpl^2/H^3$~\cite{Dvali:2017eba}. In the black hole analogy, this characteristic time scale corresponds to the Page time~\cite{Page:1993wv, Page:2013dx}, after which the semi-classical, perturbative approach can be no longer valid. The Page time in black hole physics compare the entropy of thermal radiation and that of the black hole horizon to see if black hole evaporation is consistent with unitarity. Since dS in the Unruh vacuum mimics an evaporating black hole, we expect that it also has the information paradox problem.

Meanwhile, the developments in string theory suggest that every parameter in physics except for the string length is determined dynamically through the moduli stabilization. If a scalar responsible for the dS vacuum energy, say, the inflaton, also determines the masses of some other particles, it may induce stronger backreaction that destabilizes dS~\cite{Ooguri:2018wrx} (see \cite{Seo:2018abc} for discussion on more generic inflationary scenario beyond slow-roll inflation). In the thermodynamic point of view, this becomes evident when the particle masses become light along the inflaton trajectory, which invalidates effective theory by producing rapidly increasing amount of entropy (for more discussion in terms of thermodynamics, see, e.g.,~\cite{Seo:2019mfk}). This can be another way that dS is destabilized through the increase in another entropy contributions, even if thermal radiation alone makes $\epsilon_H < 0$. It thus indicates that the degrees of freedom other than the inflaton, the isocurvature perturbations, could provide a unique window to investigate the evolution of the quasi-dS phase in the early universe -- while it may not be easy to maintain the isocurvature fields light in general field space~\cite{Gong:2011uw}.

\subsection*{Acknowledgements}

We thank Robert Brandenberger and Gungwon Kang for useful discussions. 
JG is supported in part by the National Research Foundation of Korea grant funded by the Korean Government (2019R1A2C2085023). JG also acknowledges the Korea-Japan Basic Scientific Cooperation Program supported by the National Research Foundation of Korea and the Japan Society for the Promotion of Science (2020K2A9A2A08000097). 
JG is grateful to the Asia Pacific Center for Theoretical Physics for hospitality while this work was under progress.

\newpage

\appendix

\renewcommand{\theequation}{\Alph{section}.\arabic{equation}}

\section{Details of Section~\ref{sec:Tmunu}}
\label{app:detsec3}
\setcounter{equation}{0}

\subsection{Bogoliubov transformation in detail}
\label{app:Bogoliubov}

Here we show the details of the Bogoliubov transformation considered in Section~\ref{sec:Tmunu}. Our conventions mainly follow~\cite{Birrell:1982ix}, except for the sign of the metric. The inner product of the solutions to the Klein-Gordon equation on a space-like hypersurface $\Sigma$ normal to the time direction, on which the induced spatial metric being $g_\Sigma$, is defined as
\begin{equation}
(u_i, u_j) \equiv - i \int_\Sigma d\Sigma^\mu \sqrt{|g_\Sigma|} 
\Big[ u_i \nabla_\mu u_j^* - (\nabla_\mu u_i) u_j^* \Big]
\, ,
\end{equation}
which satisfies $(u_i, u_j)=(u_j, u_i)^*=-(u_j^*, u_i^*)$. Given a scalar field $\varphi$, the annihilation (creation) operator associated with the positive frequency mode $u_i$ ($u_i^*$) is given by $a_i=(u_i, \varphi)$ [$a_i^\dagger=-(u_i^*, \varphi)$], from which the following relations follow:
\begin{equation}
\big[ a_i, a_j^\dagger \big] = (u_i, u_j)
\quad \text{and} \quad
\big[ a_i, a_j \big] = \big[ a_i^\dagger, a_j^\dagger \big] = 0 \, .
\end{equation}
When $\varphi$ is expressed in terms of two differently defined positive frequency modes $u_i$ and $\overline{u}_i$ as
\begin{equation}
\varphi(x) = \sum_i \Big[ a_i u_i(x) + a_i^\dagger u_i^* \Big]
= \sum_i \Big[ \overline{a}_i \overline{u}_i(x) + \overline{a}_i^\dagger \overline{u}_i^* \Big]
\, ,
\end{equation}
two different representations are related by the Bogoliubov transformations for the mode functions and operators respectively:
\begin{alignat}{2}
\overline{u}_i & = \sum_j \Big[ \alpha_{ij} u_j + \beta_{ij} u_j^* \Big] \, ,
\qquad
u_i & = \sum_j \Big[ \alpha_{ji}^* \overline{u}_j - \beta_{ji} \overline{u}_j^* \Big] \, ,
\\
\overline{a}_i & = \sum_j \Big[ \alpha_{ij}^* a_j - \beta_{ij}^* a_j^\dagger \Big] \, ,
\qquad
a_i & = \sum_j \Big[ \alpha_{ji} \overline{a}_j + \beta_{ji}^* \overline{a}_j^\dagger \Big] \, .
\end{alignat}
Here the coefficients $\alpha_{ij}$ and $\beta_{ij}$ are given by
\begin{equation}
\alpha_{ij} = \big( \overline{u}_i, u_j \big)
\quad \text{and} \quad 
\beta_{ij} = - \big( \overline{u}_i, u_j^* \big) \, ,
\end{equation}
which satisfy the following relations:
\begin{equation}
\sum_k \big[ \alpha_{ik} \alpha_{jk}^* - \beta_{ik} \beta_{jk}^* \big] = \delta_{ij}
\quad \text{and} \quad
\sum_k \big[ \alpha_{ik} \beta_{jk} - \beta_{ik} \alpha_{jk} \big] = 0 \, .
\end{equation}

\subsection{Normalization of positive frequency mode function}
\label{app:normalization}

We here determine the normalization constant $C_{\omega\ell}$ in the positive frequency mode function, $\varphi(t_s, r_s, \theta, \phi)=C_{\omega\ell} R_{\omega  \ell}(H r_s)Y_{\ell}^m(\theta, \phi)e^{-i\omega t_s}$.
Requiring the proper normalization
\begin{equation}
(\varphi_i, \varphi_j) = \delta_{ij} 
\quad \text{for} \quad
i,j = (\omega, \ell, m)
\, ,
\end{equation}
we obtain
\begin{align}
\label{eq:norm}
\frac{\delta(\omega-\omega')}{2\omega}
& =
|C_{\omega\ell}|^2 \int_0^{1/H} \frac{r_s^2 dr_s}{1-H^2r_s^2}
R_{\omega  \ell}(H r_s) R_{\omega'  \ell}^*(H r_s)
\nonumber\\
& =
\frac{|C_{\omega\ell}|^2}{2H^3} \int_0^1 du \frac{\sqrt{u}}{1-u} 
R_{\omega  \ell}(\sqrt{u}) R_{\omega'  \ell}^*(\sqrt{u})
\, ,
\end{align}
where we have defined $u \equiv H^2r_s^2$ in the last step. While the explicit calculation is challenging, we  expect that the integrand is dominant around $u\simeq 1$. Using \eqref{eq:horwave} for $u \simeq 1$ and
\begin{equation}
\int_0^1 du \frac{\sqrt{u}}{1-u} (1-u)^{-i\omega/(2H)}
\simeq 
- \int_0^1 d\log(1-u) e^{-i\frac{\omega}{2H} \log(1-u)}
=
\pi \delta \bigg( \frac{\omega}{2H} \bigg)
\, ,
\end{equation}
and noting that $\delta(\omega+\omega')=0$ for $\omega, \omega' >0$ we find 
\begin{equation}
\eqref{eq:norm} = \frac{|C_{\omega\ell}|^2}{2H^3} 4\pi H |A_{\omega\ell}|^2 \delta(\omega-\omega') \, ,
\end{equation}
so that $C_{\omega\ell}$ is determined as
\begin{equation}
|C_{\omega\ell}| = \frac{H}{\sqrt{4\pi \omega}|A_{\omega\ell}|} \, .
\end{equation}
Choosing an appropriate phase for $C_{\omega\ell}$, we find that near the horizon $r_s\to H^{-1}$, $\varphi$ is written as
\begin{equation}
\varphi(t_s, r_s, \theta, \phi) 
\simeq 
\frac{H}{\sqrt{4\pi\omega}}
\Big[ e^{-i\delta_{\omega\ell}} e^{-i\omega(t+r_*)} + e^{+i\delta_{\omega\ell}} e^{-i\omega(t-r_*)} \Big] Y_\ell^m(\theta,\phi) \, ,
\end{equation}
where $\delta_{\omega\ell}={\rm arg}[A_{\omega\ell}]-i(\omega/H)\log 2$.

\section{Geometric properties of the Killing horizon}
\label{app:Killhor}
\setcounter{equation}{0}

Here we briefly summarize several properties of the null hypersurface and the Killing horizon used in Section~\ref{sec:area}. For further detail, we refer the reader to, for example,~\cite{Wald:1984rg, Reall}.

A null hypersurface ${\cal N}$ can be defined by some function $f=$ constant satisfying $g_{ab}(df)^a (df)^b=0$. Since the normal vector $U^a\equiv g^{ab}(df)_b$ in this case is also tangent to ${\cal N}$, we can consider the geodesic parametrized by an affine parameter $\lambda$ such that $U^a=(\partial/\partial\lambda)^a$. Then the null geodesic congruence, i.e., a family of geodesics that passes any point on ${\cal N}$ exactly once, can be parametrized by $(\lambda, s)$, where $s$ parametrizes the geodesic deviation. From $s$ the deviation vector is defined by $S^a=(\partial/\partial s)^a$. Since $(\lambda, s)$ form the coordinate chart, we can set $[U, S]=0$ such that the change in $S^a$ along the geodesic is written as $U^b \nabla_b S^a = S^b \nabla_b U^a \equiv B^a{}_{b} S^b$, where we define $B^a{}_b\equiv\nabla_b U^a$. Note that $U^2=0$ implies that $B^a{}_b U^b=0$.

When we introduce another vector $N^a$ satisfying $N^2=0$, $N\cdot U=-1$ and $U\cdot \nabla N^a=0$, we can make a decomposition $S^a=\alpha U^a+\beta N^a+\widehat{S}^a$, where $\widehat{S}^a$ is a vector orthogonal to both $U^a$ and $N^a$ by projecting them out of $S^a$: $\widehat{S}^a \equiv P^a{}_b S^b$ where $P^a{}_b \equiv \delta^a{}_b + N^aU_b + U^a N_b$. In our case, $S^a$ is tangent to ${\cal N}$, i.e., orthogonal to the normal vector $U^a$ hence $\beta=0$. Since $U\cdot S=0$, we can define the projected $B^a{}_b$, $\widehat{B}^a{}_b=P^a{}_cB^c{}_dP^d{}_b$, such that $\widehat{S}^a$ obeys the equation in the same form as that of $S^a$: $U^b\nabla_b \widehat{S}^a=\widehat{B}^a{}_b\widehat{S}^b$. Then $\widehat{B}^a{}_b$ is decomposed into $\widehat{B}_{ab}=\Theta P_{ab}/2+\widehat{\sigma}_{ab}+\widehat{\omega}_{ab}$, where 
\begin{alignat}{2}
\text{Expansion} & :
\quad &
\Theta & = \widehat{B}^a{}_a = B^a{}_a \, ,
\\
\text{Shear} & :
\quad &
\widehat\sigma_{ab} & = \widehat{B}_{(ab)} - \frac{1}{2}P_{ab}\Theta \, ,
\\
\text{Rotation} & :
\quad &
\widehat\omega_{ab} & = \widehat{B}_{[ab]} \, .
\end{alignat}
Here, the hats for the shear and rotation indicate that they are only with respect to the transverse directions, as we are considering null geodesic congruences.

When the Killing field $\xi_a$ satisfies $\xi^2=0$ on ${\cal N}$ hence normal to ${\cal N}$, we call ${\cal N}$ the Killing horizon. The surface gravity $\kappa$ of the Killing horizon is defined by
\begin{equation}
\nabla_a (\xi_b\xi^b)|_{\cal N} = -2\xi^b\nabla_b\xi_a \equiv -2\kappa \xi_a \, ,
\end{equation}
where for the first equality the Killing equation $\nabla_a\xi_b+\nabla_b\xi_a=0$ is used. For dS, using the statement below \eqref{eq:dSkruskal}, we find that $\kappa=H$ on both ${\cal H}^-$ and ${\cal H}^+$. The first law of thermodynamics implies that the temperature of the Killing horizon is given by $T=\kappa/(2\pi)$.

\subsection{Vanishing of $\widehat{B}_{ab}$ on the Killing horizon}
\label{app:vanKill}

Here we briefly sketch the justification for the vanishing of $\widehat{B}_{ab}$ on the Killing horizon.

\begin{enumerate}

\item On the null hypersurface, $\widehat{\omega}_{ab}=0$.

Since $U^a B_{ab}=B_{ba}U^a=0$, we have $\widehat{B}^b{}_c = B^a{}_b + U^b N_d B^d{}_c + U_cB^b{}_dN^d + U^bU_cN_dB^d{}_eN^e$, then
\begin{equation}
U_{[a} \widehat{\omega}_{bc]} = U_{[a} \widehat{B}_{bc]} = U_{[a} B_{bc]} = U_{[a} \nabla_{c}U_{b]} = 0 \, ,
\end{equation}
where the last equality comes from $dU^a=0$ as implied by $U_a=(df)_a$. Rewriting the leftmost expression as $(U_{a}\widehat{\omega}_{bc} + U_{b}\widehat{\omega}_{ac} + U_{c}\widehat{\omega}_{ab})/3$ and contracting it with $N^a$, the facts $N\cdot U=-1$ and $\widehat{\omega}_{ab}N^b=0$ give $\widehat{\omega}_{ab}=0$.

\item On the Killing horizon, $\Theta$ and $\widehat{\sigma}_{ab}$ vanish.
 
Since the Killing field $\xi^a$ is normal to ${\cal N}$, we can express $U^a=h\xi^a+f V_a$, where $h$ is some smooth function, $f$ is a smooth function that vanishes on ${\cal N}$ as defined before (hence $df\sim U_a$) and $V_a$ is some smooth vector field. Then after symmetrizing
\begin{equation}
B_{ab} = \nabla_b U_a = (\nabla_b h)\xi_a + h\nabla_b \xi_a + (\nabla_b f)V_a + f\nabla_b V_a \, ,
\end{equation}
and using the fact that $f=0$ on the horizon and the Killing equation $\nabla_{(a}\xi_{b)}=0$, we obtain 
\begin{equation}
B_{(ab)}|_{\cal N} = \big( \xi_{(a}\nabla_{b)} h + V_{(a}\nabla_{b)} f \big) |_{\cal N} 
= h^{-1} U_{(a} \nabla_{b)} h + V_{(a} U_{b)} \, ,
\end{equation}
which leads to $\widehat{B}_{(ab)} = P_{ac} B^{c}{}_d P^d{}_b = 0$ by $P^a{}_b U^b = 0$. Then both $\Theta$ and $\widehat{\sigma}_{ab}$ vanish.

\end{enumerate}

\subsection{Area determined by expansion}
\label{app:area}

The expansion $\Theta$ determines the area of the null hypersurface along $\lambda$. This can be easily seen in the Gaussian null coordinates, in which an affine parameter $r$ is introduced such that ${\cal N}$ corresponds to $r=0$ and $V^a=(\partial/\partial r)^a$ satisfies $V^2=0$, $V\cdot U=1$, and $V\cdot(\partial/\partial y^i)=0$, with $y^i$ being two-dimensional coordinates on ${\cal N}$ other than $\lambda$. Then the metric on ${\cal N}$ is written as $ds^2 = 2 dr d\lambda + h_{ij} dy^i dy^j$. In $(r, \lambda, y^i)$ coordinates, the components of $U^a$ are given by $U^\mu=(0,1,0,0)$ or $U_\mu=(1,0,0,0)$. Then the equalities $U^a B_{ab} = B_{ba} U^a = 0$ become the conditions $B^r{}_\mu = B^\mu{}_\lambda = 0$. From these we obtain
\begin{equation}
\Theta = B^a{}_a = \nabla_a U^a = \nabla_i U^i = \Gamma^i_{i\lambda} 
= \frac12 h^{ij} h_{ij,\lambda} = \frac{1}{\sqrt{h}} \partial_\lambda \sqrt{h} \, ,
\end{equation}
or
\begin{equation}
\frac{d}{d \lambda}{\cal A} = \frac{d}{d \lambda} \int \sqrt{h} = \int \Theta \sqrt{h} = \int \Theta d{\cal A} \, .
\end{equation}

\subsection{The Raychaudhuri equation}
\label{app:Rayeq}

The change in the expansion $\Theta$ along $\lambda$ is determined by the Raychaudhuri equation. Since both $U$ and $N$ are parallel transported along $\lambda$, i.e., $U^b \nabla_b U_a = U^b \nabla_b N_a = 0$, we know $U \cdot \nabla P^a{}_b = 0$, from which we obtain
\begin{align}
\frac{d\Theta}{d\lambda}
& =
U \cdot \nabla \big( B^a{}_b P^b{}_a \big)
=
P_a{}^b U \cdot \nabla B^a{}_ b 
=
P_a{}^b U^c \big( \nabla_b \nabla_c U^a + R^a{}_{dcb} U^d \big)
\nonumber\\
& =
P_a{}^b \big[ - (\nabla_b U^c) (\nabla_c U^a) + R^a{}_{dcb} U^c U^d \big]
\nonumber\\
& =
- B_b{}^c P_a{}^b B^a{}_c - R_{cd} U^c U^d
\nonumber\\
& =
- \widehat{B}_b{}^c \widehat{B}^a{}_c - R_{cd} U^c U^d
\, ,
\end{align}
hence the Raychaudhuri equation follows:
\begin{equation}
\label{eq:raychaudhuri}
\frac{d\Theta}{d\lambda}
= 
- \frac12 \Theta^2 - \widehat{\sigma}^{ab}\widehat{\sigma}_{ab} 
+ \widehat{\omega}^{ab}\widehat{\omega}_{ab} - R_{cd} U^c U^d
\, .
\end{equation}

\end{document}